\PassOptionsToPackage{unicode}{hyperref}
\PassOptionsToPackage{hyphens}{url}
\PassOptionsToPackage{dvipsnames,svgnames,x11names}{xcolor}
\documentclass[
  12pt]{article}

\usepackage{booktabs}
\usepackage{setspace}
\usepackage{amsmath,amssymb}
\usepackage{iftex}
\ifPDFTeX
  \usepackage[T1]{fontenc}
  \usepackage[utf8]{inputenc}
  \usepackage{textcomp} 
\else 
  \usepackage{unicode-math}
  \defaultfontfeatures{Scale=MatchLowercase}
  \defaultfontfeatures[\rmfamily]{Ligatures=TeX,Scale=1}
\fi
\usepackage{lmodern}
\ifPDFTeX\else  
\fi
\IfFileExists{upquote.sty}{\usepackage{upquote}}{}
\IfFileExists{microtype.sty}{
  \usepackage[]{microtype}
  \UseMicrotypeSet[protrusion]{basicmath} 
}{}
\makeatletter
\usepackage{xcolor}
\makeatletter
\ifx\paragraph\undefined\else
  \let\oldparagraph\paragraph
  \renewcommand{\paragraph}{
    \@ifstar
      \xxxParagraphStar
      \xxxParagraphNoStar
  }
  \newcommand{\xxxParagraphStar}[1]{\oldparagraph*{#1}\mbox{}}
  \newcommand{\xxxParagraphNoStar}[1]{\oldparagraph{#1}\mbox{}}
\fi
\ifx\subparagraph\undefined\else
  \let\oldsubparagraph\subparagraph
  \renewcommand{\subparagraph}{
    \@ifstar
      \xxxSubParagraphStar
      \xxxSubParagraphNoStar
  }
  \newcommand{\xxxSubParagraphStar}[1]{\oldsubparagraph*{#1}\mbox{}}
  \newcommand{\xxxSubParagraphNoStar}[1]{\oldsubparagraph{#1}\mbox{}}
\fi
\makeatother

\usepackage{longtable,booktabs,array}
\usepackage{calc} 
\usepackage{etoolbox}
\makeatletter
\patchcmd\longtable{\par}{\if@noskipsec\mbox{}\fi\par}{}{}
\makeatother
\IfFileExists{footnotehyper.sty}{\usepackage{footnotehyper}}{\usepackage{footnote}}
\makesavenoteenv{longtable}
\usepackage{graphicx}
\makeatletter
\def\maxwidth{\ifdim\Gin@nat@width>\linewidth\linewidth\else\Gin@nat@width\fi}
\def\maxheight{\ifdim\Gin@nat@height>\textheight\textheight\else\Gin@nat@height\fi}
\makeatother
\setkeys{Gin}{width=\maxwidth,height=\maxheight,keepaspectratio}
\makeatletter
\def\fps@figure{htbp}
\makeatother

\usepackage[subtle]{savetrees}
\usepackage{amsmath}
\usepackage{graphicx}
\usepackage{enumerate}
\usepackage{natbib}
\usepackage{url} 
\usepackage{float}
\usepackage{ragged2e}
\usepackage{amsthm}
\usepackage{mathtools}
\usepackage{amsfonts}
\usepackage{placeins}
\usepackage{makecell}
\usepackage{amssymb}
\usepackage{hyperref}  
\usepackage{caption}
\usepackage{subcaption}
\usepackage{apptools}
\usepackage{xcolor}
\usepackage{xr}
\usepackage{adjustbox}
\usepackage{wrapfig}
\usepackage{caption}
\usepackage{subcaption}
\usepackage{cprotect}
\usepackage{enumitem}
\usepackage{array} 
\usepackage{placeins}
\usepackage{bbm}
\usepackage{multirow}
\usepackage{dsfont}
\usepackage{xcolor}
\usepackage{tabularx}
\input{0_commands}
\theoremstyle{definition}

\newtheorem{theorem}{Theorem}
\newtheorem{assumption}{Assumption}
\newtheorem{algorithm}{Algorithm}

\newtheorem{definition}{Definition}
\newtheorem{remark}{Remark}
\newtheorem{corollary}{Corollary}
\newtheorem{lemma}{Lemma}
\newtheorem{example}{Example}

\AtAppendix{\counterwithin{theorem}{section}}
\AtAppendix{\counterwithin{assumption}{section}}
\AtAppendix{\counterwithin{algorithm}{section}}
\AtAppendix{\counterwithin{proposition}{section}}
\AtAppendix{\counterwithin{definition}{section}}
\AtAppendix{\counterwithin{remark}{section}}
\AtAppendix{\counterwithin{corollary}{section}}
\AtAppendix{\counterwithin{lemma}{section}}
\AtAppendix{\counterwithin{example}{section}}

\newcommand{\anon}{1}

\begin{document}

\def\spacingset#1{\renewcommand{\baselinestretch}%
{#1}\small\normalsize} \spacingset{1}

\if1\anon
{
  \title{\bf Automatic debiased machine learning \\for dynamic treatment effects \\and general nested functionals}
 \author{
\begin{tabular}{cc}
Victor Chernozhukov & Whitney K. Newey \\
MIT Economics & MIT Economics
\\[1em]
Rahul Singh\thanks{We thank seminar participants at ESIF 2024, NYU, BU, Simon Fraser, UBC, Cornell, and Texas A\&M. We thank Raj Chetty for sharing data for the empirical application. Juhui Jin, Marvin Lob, Arjun Nageswaran, Jinho Park, Wisse Rutgers, and Joshua Zhang provided excellent research assistance.}
& Vasilis Syrgkanis \\
Harvard Economics & Stanford MS\&E
\end{tabular}
}

\date{First draft: March 2022 \\ This draft: July 2026}

  \maketitle
} \fi

\if0\anon
{
  \bigskip
  \bigskip
  \bigskip
  \begin{center}
    {\LARGE\bf Automatic Debiased Machine Learning for Dynamic Confounding and General Nested Functionals}
\end{center}
  \medskip
} \fi

\bigskip

\begin{abstract}
Many canonical models in causal inference and structural econometrics have recursive identification formulas.
In causal inference, recursion arises when identification requires both pre- and post-treatment covariates.
For example, short-term surrogate outcomes are measured after the treatment, and serve as necessary covariates when identifying long-term effects.
Post-treatment covariates are also required for identification of dynamic difference-in-differences designs, time-varying treatment regimes, and mediation analysis.
In structural econometrics, recursion arises through evolving state variables, for example in dynamic sample selection models and dynamic discrete choice models.
In this paper, we propose an automatic and recursive method for inference, applicable to such formulas, allowing for flexible estimation by neural networks and random forests. As a technical contribution, we introduce recursive Riesz representers. 
\end{abstract}

\noindent%
\vfill

\newpage
\spacingset{1.5} 

\setlength{\abovedisplayskip}{6pt plus 2pt minus 2pt}
\setlength{\belowdisplayskip}{6pt plus 2pt minus 2pt}
\setlength{\abovedisplayshortskip}{3pt plus 1pt minus 1pt}
\setlength{\belowdisplayshortskip}{3pt plus 1pt minus 1pt}
\setlength{\jot}{3pt}

\section{Introduction}\label{sec:intro}

Several canonical models in causal and structural econometrics have a recursive identification formula: the parameter is identified as a functional, or scalar summary, of a function that is recursively defined. Causal examples include the surrogacy \citep{prentice1989surrogate,athey2025surrogate}, dynamic difference-in-differences \citep{caetano2022difference}, time-varying treatment \citep{robins1986new}, and mediation \citep{pearl2001direct,imai2010identification} models. Structural examples include dynamic sample selection \citep{robins1995analysis,abowd2001moment} and dynamic discrete choice models  \citep{rust1987optimal,heckman2007dynamic}. 
These examples are widely used in empirical economic research; 
among all papers published in ``top five'' journals between $2010$ and $2025$, $14\%$ have recursive elements that fit within our framework.\footnote{Out of the $5484$ articles, $306$ use explicit recursive formulas, while an additional $452$ use recursive identifying assumptions in the primary or supporting analyses.}

As a concrete example, suppose we wish to estimate a long-term effect by combining short-term experimental data with long-term observational data. The experimental sample contains baseline covariates $X$, the conditionally randomized treatment $D$, and short-term surrogate outcomes $S$. The observational sample contains $(X,S)$ and the long term outcome of interest $Y$. The identification formula for the average treatment effect is recursive: $f_2(X,S)=\E_{\obs}(Y|X,S)$ is the surrogate index identified by observational data, $f_1(X,D)=\E_{\exp}\{f_2(X,S)|X,D\}$ is the conditional average response in the experimental sample, and $\theta_0=\E\{f_1(X,1)-f_1(X,0)\}$ is the average treatment effect for the entire population \citep{athey2025surrogate}. Formally, the surrogate formula for the treatment effect $\theta_0$ is a functional of $f_1$, which is a short term conditional expectation of $f_2$, which is itself a long term conditional expectation. Intuitively, identification involves recursively expecting the future.

More generally, identification formulas are recursive when the model has three ingredients. 
The first ingredient is short panel data: it must follow a growing cohort of units over a fixed number of time periods. In surrogacy, the time periods are ``short-term'' and ``long-term'', but there can be more than two periods. 
The second ingredient is what we call ``dynamics on observables'': after unit-level conditioning at each time period, the remaining transition dynamics are as good as random. In the surrogacy model, treatment assignment is as good as random after conditioning on pre-treatment covariates, and the surrogate index can be transported from one sample  to the other after conditioning on the pre- and post-treatment covariates. 
The third ingredient is nested conditioning: the covariate histories in earlier time periods are nested by those in later time periods. Nonetheless, changes of measure across time periods are allowed, e.g. we can change from the experimental measure in the short term to the observational measure in the long term.

This paper's primary contribution is a method to automatically and recursively construct Neyman orthogonal estimating equations for general recursive functionals. 
The method is automatic in that it does not require case-by-case analytic derivations of influence functions.
In particular, it applies to an important setting where, to our knowledge, the orthogonal estimating equation has not been derived (Example~\ref{ex:diff}).
It is recursive in an intriguing way: while the known identifying equations involve recursive conditional expectations of the future, we prove that their debiasing equations involve recursive Riesz representers of the past.
After constructing the Neyman orthogonal estimating equations, we verify that the corresponding estimator is asymptotically normal at the rate $n^{-1/2}$, under product rate conditions that tolerate machine learning.
Consequently, recursively defined Riesz representers provide a unifying perspective on the asymptotic variance of semiparametric estimators for recursive functionals.

As a secondary contribution, we propose and analyze an estimator for the recursive Riesz representer. 
The estimator is defined over general function spaces, such as neural networks and random forests. 
It amounts to new, neural network-based balancing weights for many causal and structural models. 
We derive a nonasymptotic bound on its estimation error, which satisfies the rate condition needed for semiparametric inference. 
The technical challenge relative to previous work is to simultaneously (i) account for the accumulation of estimation error when recursing across time periods, and (ii) allow for general machine learning. 
In simulations, our method reduces bias and improves precision relative to some previous methods, because our flexible approach greatly reduces the approximation error. 
In real-world applications, it allows for flexible causal modeling without sacrificing statistical power.

\subsection{Related work}

The vast majority of literature on machine learning-based balancing weights and Riesz representers is for cross sectional models, where the functionals are not recursive. Balancing weights are vectors defined in-sample; see, e.g. \cite{zubizarreta2015stable,athey2018approximate,hirshberg2021augmented} and many references therein. Balancing weights may be viewed as the in-sample evaluations of Riesz representers \citep{chernozhukov2022automatic,chernozhukov2026adversarial,chernozhukov2021automatic}, which are functions that appear in asymptotic variances \citep{newey1994asymptotic} and that can be evaluated on held-out data, e.g. in targeted \citep{zheng2011cross,van2018targeted} and debiased \citep{chernozhukov2018double,chernozhukov2023simple} machine learning. Our contribution is to generalize the nonparametric Riesz representer perspective from cross-sectional models to some short panel models, illuminating their recursive structure.

Relative to some earlier works on recursive functionals, we (i) provide a fully nonparametric derivation of the Neyman orthogonal estimating equation in terms of Riesz representers, and (ii) derive valid inference with general machine learning for Riesz representers, such as neural networks.
Our construction may be viewed as the nonparametric generalization of the recursive but parametric construction of \cite{bang2005doubly}; see Appendix~\ref{sec:clever}.
Whereas the series results of \cite{ai2003efficient} apply to a more general class of functionals than ours, their estimating equation is not Neyman orthogonal, and therefore their inference result imposes the Donsker condition, ruling out machine learning.
Finally, \cite{molina2017multiple,luedtke2017sequential,rotnitzky2017multiply} propose methods for the time-varying treatment effect, and related factorized likelihood models, which are Neyman orthogonal and can accommodate machine learning. Unlike these works, which rely on analytically derived influence functions, we construct the debiasing equations automatically through recursive Riesz representers. Our recursive Riesz estimator bypasses the issue of numerically unstable inverse propensity weights, which arise multiplicatively in their methods; see Section~\ref{sec:empirics}.

Some previous works study balancing weights for the specific functional of the time-varying treatment effect, and the specific function class of the reproducing kernel Hilbert space \citep{kallus2021optimal} and the lasso \citep{viviano2026dynamic}. Our results are complementary by (i) studying Riesz representers for a general class of recursive functionals such as the surrogate formula (even accomodating changes of measure), (ii) allowing general function classes such as neural networks, and (iii) avoiding unnecessary restrictions on potential outcomes. Neither of our main results is contained in prior work; see Sections~\ref{sec:orthogonality} and~\ref{sec:riesz} for detailed comparisons.

Broadly speaking, this paper shares the same goal as a literature that seeks to automate inference for complex causal problems; see, e.g. \cite{carone2019toward,jordan2022empirical} and many references therein.
However, our construction via recursive Riesz representers appears to be altogether different. In particular, we demonstrate that recursive structure makes seemingly complex inference problems highly tractable.

The class of recursive short panel models we study have dynamics on observables, and therefore exclude dynamic panel models with latent fixed effects \citep{nickell1981biases,arellano1991some}.  Future work may extend our construction to this important class. Nonetheless, this paper's results apply to a several causal and structural models that are extensively used in empirical economic research; over $300$ papers in ``top five'' journals from $2010$ to $2025$ use recursive models with dynamics on observables as their main empirical approach.

Section~\ref{sec:examples} defines the general class of recursive functionals, and illustrates several leading examples from econometrics and statistics.
Section~\ref{sec:orthogonality} proves our main result: automatic and recursive Neyman orthogonality for general recursive functionals. The key innovation is a recursive Riesz representer.
After defining the recursive Riesz representer in Section~\ref{sec:orthogonality}, we propose an automatic and recursive estimator for it in Section~\ref{sec:riesz}, and bound its estimation error. 
Section~\ref{sec:empirics} demonstrates the practical consequences of our theoretical contributions. In simulations, our method reduces bias and increases precision relative to some earlier methods. Our method  sheds new light on the long term effects of job training identified with surrogate outcomes, and the dynamic effects of minimum wage rules identified by dynamic difference-in-differences.
Section~\ref{sec:conc} concludes. Appendices~\ref{sec:nonlinear} and~\ref{sec:instrument} extend our Neyman orthogonal construction to nonlinear models and instrumental variable models, respectively.
\section{Examples of recursive identification}\label{sec:examples}

To motivate our main results, we demonstrate that recursive functionals proliferate across economic and statistical research. In ``top five'' journals from 2010-2025, $306$ articles use explicit recursive formulas, while an additional $452$ use recursive identifying assumptions in their analyses.

\subsection{Core econometric models}

To begin, we recap the surrogate formula from Section~\ref{sec:intro}. This family of empirical methods is widely used for estimating long term effects in economics.\footnote{From $2010$ through $2025$, we found $21$ ``top five'' articles placing surrogacy assumptions, often in labor and public economics \citep{heckman2017quantifying,hendren2020unified,dal2021information,dynarski2021closing}.}

\begin{example}[Surrogate formula]\label{ex:surrogate}
    Suppose we observe an experimental sample with baseline covariates $X$, conditionally randomized treatment $D$, and surrogate outcomes $S$, as well as an observational sample with baseline covariates $X$, surrogate outcomes $S$, and the true outcome $Y$. Under surrogacy assumptions \citep{athey2025surrogate}, the average treatment effect is identified by
    $$
    \theta_0=\E\{f_1(X,1)-f_1(X,0)\},\quad f_1(X,D)=\E_{\exp}\{f_2(X,S)|X,D\},\quad f_2(X,S)=\E_{\obs}(Y|X,S)
    $$
    where $\E_{\exp}(\cdot)$ and $\E_{\obs}(\cdot)$ are expectations over experimental and observational samples.
\end{example}

Our results also apply to some closely related formulas for data combination, which correct for a type of panel attrition with refreshment samples \citep{hirano2001combining}, or which correct for  observational selection with experimental samples \citep{athey2025experimental}. 

Among recursive functionals, one of the most widely used example in economics is dynamic sample selection. A  popular version of sample selection across multiple time periods is monotonic outcome attrition: units selectively drop out at each time period, without returning.\footnote{We found $343$ ``top five'' articles from $2010$ through $2025$ with a version of this example, of which $89$ use an explicit, recursive formula. Among the $470$ experimental ``top five'' papers we found, $109$ discussed or tested for dynamic attrition, of which $30$ explicitly adjusted for it.}

\begin{example}[Dynamic sample selection]\label{ex:sample}
    Define the time-varying covariates $X_t$ and the time-varying outcome $Y_t$ at each period $t\in\{1,2,3\}$. We observe $(X_1,X_2,X_3,Y_1)$ for all units, but we only observe $Y_2$ for a selected sample called ``wave two'', and we only observe $Y_3$ for a further selected sample called ``wave three''. If selection is monotone and as good as random conditional upon observed histories \citep{robins1995analysis,abowd2001moment}, the final outcome mean is identified by $ \theta_0=\E\{f_1(X_1,Y_1,X_2)\}$, where
    \begin{align*} 
  f_1(X_1,Y_1,X_2)&=\E_{\operatorname{wave-two}}\{f_2(X_1,Y_1,X_2,Y_2,X_3)|X_1,Y_1,X_2\}\\
  f_2(X_1,Y_1,X_2,Y_2,X_3)&=\E_{\operatorname{wave-three}}(Y_3|X_1,Y_1,X_2,Y_2,X_3).
 \end{align*}
     Here, $\E_{\operatorname{wave-two}}(\cdot)$ and $\E_{\operatorname{wave-three}}(\cdot)$ are expectations over the different waves.
\end{example}

We also contribute to a rapidly growing literature on difference-in-difference designs. An important version of the model uses time-varying covariates for identification by parallel trends. In terms of published papers, it seems the methodological work has outpaced the empirical work, but anecdotally there seem  to be many empirical papers in progress.

\begin{example}[Dynamic difference-in-differences]\label{ex:diff}
    Both before and after the treatment $D$, suppose we observe covariates $(X_{\pre},X_{\post})$ and outcomes $(Y_{\pre},Y_{\post})$. To lighten notation, let $Y_{\Delta}=Y_{\post}-Y_{\pre}$. Under dynamic parallel trends assumptions \citep{caetano2022difference}, the average treatment effect on the treated is identified by $\theta_0=\E_{\treated}(Y_{\Delta})-\E_{\treated}\{f_1(X_{\pre},Y_{\pre})\}$, where 
    $$
   f_1(X_{\pre},Y_{\pre})=\E_{\untreated}\{f_2(X_{\pre},X_{\post})|X_{\pre},Y_{\pre}\},\quad f_2(X_{\pre},X_{\post})=\E_{\untreated}(Y_{\Delta}|X_{\pre},X_{\post}).
    $$
    Here, $\E_{\treated}(\cdot)$ and $\E_{\untreated}(\cdot)$ are expectations over the treated and untreated groups.
\end{example}

In summary, recursive functionals arise in data fusion, in dynamic settings with selective missingness, and in short panels with parallel trends.

\subsection{Core statistical models}

Next, we turn to core models in statistics, which are also reasonably popular in empirical economic research. The first such example uses recursive unconfoundedness to identify the effect of a treatment sequence.\footnote{
This identification strategy appears in $13$ ``top five'' papers from 2010 through 2025, with a few papers \citep{osikominu2013quick,van2022long} extending Robins' seminal $g$-formula \citep{robins1986new}.
}

\begin{example}[Time-varying treatment effect]\label{ex:time}
    Suppose we observe time-varying covariates $X_t$ and the time-varying treatment $D_t$ at each period $t\in\{1,2,3\}$, as well as a final outcome $Y$. Under recursive unconfoundedness assumptions \citep{robins1986new}, the effect of the treatment sequence $(D_1=1,D_2=0,D_3=0)$ relative to $(D_1=0,D_2=0,D_3=0)$ is identified by 
\begin{align*}
    &\theta_0=\E\{f_1(X_1,1)-f_1(X_1,0)\},\quad  f_1(X_1,D_1)=\E\{f_2(X_1,D_1,X_2,0)|X_1,D_1\} \\
    &f_2(X_1,D_1,X_2,D_2)=\E\{f_3(X_1,D_1,X_2,D_2,X_3,0)|X_1,D_1,X_2,D_2\}\\
    &f_3(X_1,D_1,X_2,D_2,X_3,D_3)=\E(Y|X_1,D_1,X_2,D_2,X_3,D_3).
\end{align*}
\end{example}

While we illustrate the time-varying treatment effect model with three time periods, it extends to any finite number of time periods, as do our results.

Another central task in causal statistics is mediation analysis, which seeks to discern the mechanism through which a treatment affects an outcome.\footnote{ 
We found $146$ ``top five'' papers from $2010$ through $2025$ that empirically study causal mechanisms, of which $23$ explicitly use Pearl's mediation formula \citep{pearl2001direct}. 
}
Traditionally, such methods have measured how skill formation mediates the effect of early childhood interventions \citep{heckman2013understanding, gronqvist2020understanding, kosse2020formation, berger2025impact}. Recent work conducts mediation analysis  more broadly \citep{abebe2021anonymity, schwardmann2022self, ang2023birth, deserranno2025allocation}. 

\begin{example}[Mediation analysis]\label{ex:mediation}
    Suppose we observe baseline covariates $X$, conditionally randomized treatment $D$, a possible  mediator $M$, and the outcome $Y$. Under mediation assumptions \citep{imai2010identification}, the natural indirect effect (i.e. the effect of $D$ on $Y$ mediated by $M$) is identified by $\theta_0=\E_{\treated}(Y)-\E_{\treated}\{f_1(X)\},$ where
    $$
   f_1(X)=\E_{\untreated}\{f_2(X,M)|X\},\quad f_2(X,M)=\E_{\treated}(Y|X,M).
    $$
     Here, $\E_{\treated}(\cdot)$ and $\E_{\untreated}(\cdot)$ are expectations over the treated and untreated groups.
\end{example}

Finally, we describe a classic example in computer science that has recently appeared in the study of job training and online platforms \citep{glynn2018front,katz2025digital}. In cross-sectional causal inference, unconfoundedness is closely related to the assumption that pre-treatment covariates satisfy a back-door criterion. There is also a front-door criterion, placed on pre- and post-treatment covariates, which can identify the average treatment effect. 

\begin{example}[Front door criterion]\label{ex:front}
    Suppose we observe the same variables as Example~\ref{ex:mediation}, but instead of conditional randomization, suppose the front door criterion \citep{pearl1995causal} is satisfied.  Then the average treatment effect is identified by $\theta_0=\E\{f_1(X,1)-f_1(X,0)\}$, where
$$f_1(X,D)=\E\{f_2(X,M)|X,D\},\quad f_2(X,M)=\tilde{\E}\{f_3(X,D,M)|X,M\},\quad f_3(X,D,M)=\E(Y|X,D,M)
$$
and $\tilde{\E}$ is a measure satisfying $\tilde{\E}(D|X,M)=\tilde{\E}(D|X)=\E(D|X)$.
\end{example}

In summary, recursive functionals arise in the study of treatment sequences, causal mechanisms, and generic post-treatment covariates.

\subsection{Previous work: Recursively expect the future}

The economic and statistical examples above clearly share a recursive structure. We now formalize that common structure with general notation.

\begin{definition}[Recursive functional]\label{def:linear}
    Let $V$ concatenate all observed variables across time periods. Let $T\geq 1$ be the fixed number of time periods. For each time period $t$, let $H_t$ be the time-varying conditioning variables, let $\E_{t}(\cdot)$ be the expectation with respect to the time-varying measure, and let $m_t(V,f_t)$ be a time-varying formula that is linear in $f_t$. We study recursive functionals of the form
    $$
    \theta_0=\E_0\{m_1(V,f_1)\},\quad f_t(H_t)=\E_t\{m_{t+1}(V,f_{t+1})|H_t\},\quad t=1,...,T-1,\quad f_T(H_T)=\E_T(Y|H_T).
    $$
\end{definition}

The class of functionals in Definition~\ref{def:linear} is recursive in that, for each time period $t$, the function $f_t$ is recursively defined from the function $f_{t+1}$. In particular, $f_t$ recursively expects the future $f_{t+1}$ using information $H_t$. As demonstrated above, several commonly used models are identified in this way. One can read off the number of time periods $T$ and the time-varying objects $\{H_t,\E_t(\cdot),m_t(\cdot)\}^{T}_{t=1}$ from the many economic and statistical examples above.

Concretely, for Example~\ref{ex:surrogate}, the concatenation of variables is $V=(X,D,S,Y)$. The short term and long term give $T=2$ time periods. The base expectation is for the full population: $\E_0(\cdot)=\E(\cdot)$. For period $t=1$, we have the conditioning variables $H_1=(X,D)$ and the expectation $\E_1(\cdot)=\E_{\exp}(\cdot)$. The formula applied to $f_1$ is $m_1(V,f_1)=f_1(X,1)-f_1(X,0)$. For period $t=2$, we have the conditioning variables $H_2=(X,S)$ and the expectation $\E_2(\cdot)=\E_{\obs}(\cdot)$. The formula applied to $f_2$ is $m_2(V,f_2)=f_2(X,S)$.

In causal and structural econometrics, an important set of parameters involve derivatives with respect to continuous interventions; see, e.g., \citet{altonji2005cross,singh2025sequential}. These examples also satisfy Definition~\ref{def:linear}, with $m_T(V,f_T)$ taken to be the derivative of $f_T$ with respect to the intervention.\footnote{The corresponding Riesz representer $a_T$ contains the score function, which our framework directly accommodates. The argument follows from integration by parts.}

While Definition~\ref{def:linear} focuses on recursive functionals with linear time-varying formulas $m_t(\cdot)$, our results extend to recursive functionals  with nonlinear time-varying formulas. Appendix~\ref{sec:nonlinear} develops this extension. The extension is highly practical, since it allows our framework to handle some dynamic discrete choice models.\footnote{We found $235$ ``top five'' papers from 2010 through 2025 using dynamic structural models, of which  $188$ use dynamic discrete choice.} Common applications are labor \citep{eckstein2011dynamic, kennan2011effect, lise2011consumption,heckman2018returns,llull2018immigration,traiberman2019occupations,arcidiacono2025college}, trade \citep{caliendo2019trade, brancaccio2020geography}, and industrial organization \citep{lee2013vertical, huang2014dynamic, choo2015dynamic, arcidiacono2016estimation, igami2020mergers,  williams2022welfare,  verdier2022welfare, fan2025estimating}. As such, our main results apply to both causal and structural models.

\section{Automatic and recursive orthogonality}\label{sec:orthogonality}

Our main contribution is a procedure to construct Neyman orthogonal estimating equations for recursive functionals. Whereas existing identification formulas recursively expect the future, our construction recursively represents the past. It is automatic in that it does not require analytic derivation of influence functions. It applies to an important setting where, to our knowledge, the orthogonal estimating equation has not been derived (Example~\ref{ex:diff}).

After constructing the general Neyman orthogonal estimating equation, we confirm that it allows for flexible machine learning, correctly adjusts for dynamic confounding and dynamic biases, and leads to a valid confidence interval for the recursively identified parameter of interest. Our confidence interval for recursively identified parameters shrinks at same $n^{-1/2}$ rate as a confidence interval for a simple, cross sectional causal parameter.

\subsection{Intuition: Long term treatment effect}

For intuition, we walk through our derivation for the average treatment effect on a long term outcome, identified by the surrogate formula. We use the notation of Example~\ref{ex:surrogate}.

To begin, focus on the short term variables $(X,D,S)$. They appear in the equations
$$
\theta_0=\E\{f_1(X,1)-f_1(X,0)\},\quad f_1(X,D)=\E_{\exp}\{f_2(X,S)|X,D\}.
$$
Viewing $f_2(X,S)$ as a pseudo-outcome, this problem resembles the cross-sectional average treatment effect with covariate shift: the regression $f_1$ is estimated from the experimental sample, which has an experimental covariate distribution, while the treatment effect $\theta_0$ is estimated from the full sample, which has a possibly different covariate distribution. 

Similar to the cross-sectional case, there is a dual expression for the treatment effect:
$$
\theta_0=\E_{\exp}\{a_1(X,D)f_1(X,D)\},\quad a_1(X,D)=\frac{\density(X)}{\density_{\exp}(X)}\left\{\frac{D}{\E_{\exp}(D|X)}-\frac{1-D}{1-\E_{\exp}(D|X)}\right\},
$$
where $\density(X)$ is the covariate density in the full population, and $\density_{\exp}(X)$ is the covariate density in the experimental population. The derivation is immediate from the law of iterated expectations. Here, $a_1(X,D)$ may be viewed as generalized balancing weights; formally, it is called a Riesz representer. Within $a_1(X,D)$, the second factor contains the familiar inverse propensity weights. The first factor is a density ratio to adjust for covariate shift. Intuitively,  $a_1(X,D)$ balances the covariates of the treated and untreated experimental subpopulations, and balances the covariates of the experimental and full populations.

We can begin to debias in the same way we would for the cross-sectional case. Combining both expressions for the treatment effect,
$$
\theta_0=\E\{f_1(X,1)-f_1(X,0)\}+\E_{\exp}[a_1(X,D)\{f_2(X,S)-f_1(X,D)\}],
$$
where the second term is zero by the law of iterated expectations. Until this point, the derivation is from earlier work on cross-sectional models \citep{chernozhukov2023automatic}. However, the derivation  is fundamentally incomplete: the estimating equation above is orthogonal in $f_1$ but not $f_2$, so we are not done.

The key insight of this paper is that we can continue to orthogonalize in a recursive manner. Now, we turn our attention to the long term variables $(X,S,Y)$. Appealing to the definition of $a_1$ as a Riesz representer, some terms in the previous equation cancel, and we arrive at
$$
\theta_0=\E_{\exp}\{a_1(X,D)f_2(X,S)\},\quad f_2(X,S)=\E_{\obs}(Y|X,S).
$$
Viewing $f_2$ as a regression and $\theta_0$ as its scalar summary, once again we can use the law of iterated expectations to derive a dual expression in terms of a new Riesz representer $a_2$:
$$
\theta_0=\E_{\obs}\{a_2(X,S)f_2(X,S)\},\quad a_2(X,S)=\frac{\density_{\exp}(X,S)}{\density_{\obs}(X,S)}\E_{\exp}\{a_1(X,D)|X,S\}
$$
where $\density_{\exp}(X,S)$ is a joint density in the experimental population, and $\density_{\obs}(X,S)$ is a joint density in the observational population. Intuitively, $a_2(X,S)$ may be viewed as weights which balance the covariates and surrogates of the experimental and observational populations. 

We conclude that the Neyman orthogonal estimating equation is
$$
\theta_0=\underbrace{\E\{f_1(X,1)-f_1(X,0)\}}_{\text{identification}}+\underbrace{\E_{\exp}[a_1(X,D)\{f_2(X,S)-f_1(X,D)\}]}_{\text{base case debiasing}}+\underbrace{\E_{\obs}[a_2(X,S)\{Y-f_2(X,S)\}]}_{\text{recursive debiasing}}
$$
where the second and third terms are zero by the law of iterated expectations. Importantly, the regressions recurse forward in time, in that $f_1$ is defined from $f_2$:
$$
f_1(X,D)=\E_{\exp}\{f_2(X,S)|X,D\}.
$$
Meanwhile, the Riesz representers recurse backwards in time, in that $a_2$ is defined from $a_1$:
$$
\E_{\obs}\{a_2(X,S)f_2(X,S)\}=\E_{\exp}\{a_1(X,D)f_2(X,S)\}.
$$
While the expressions for $a_1$ and $a_2$ seem complex, we propose an automatic and recursive method for deriving these functions below.

Moreover, we propose an automatic and recursive method for estimating $a_1$ and $a_2$ in Section~\ref{sec:riesz}. Our method bypasses the numerically unstable steps of estimating and inverting the propensity score, and covariate density, and surrogate density: $\E_{\exp}(D|X)^{-1}$, $\density_{\exp}(X)^{-1}$, and $\density_{\obs}(X,S)^{-1}$ are challenging to estimate when $X$ has many dimensions.

\subsection{This work: Recursively represent the past}

The same technique applied to Example~\ref{ex:surrogate} applies to the entire class of functionals in Definition~\ref{def:linear}. The key regularity condition is that the recursive Riesz representers exist, which we now formalize. Our condition is essentially a generalization of the standard overlap condition.

\begin{assumption}[Existence]\label{assumption:existence}
    Consider the general notation of Definition~\ref{def:linear}. For each time period $t=1,...,T$, there exists a Riesz representer $a_t(H_t)$ with finite variance satisfying
    $$
    \E_t\{a_t(H_t)\tilde{f}_t(H_t)\}=\E_{t-1}\{a_{t-1}(H_{t-1})m_t(V,\tilde{f}_t)\}
    $$
    for all placeholder functions $\tilde{f}_t(H_t)$ with finite variance. As a convention, we take $a_0=1$.
\end{assumption}

In cross sectional causal inference, $T=1$ and Assumption~\ref{assumption:existence} collapses down to the classic overlap conditions. Concretely, in Example~\ref{ex:surrogate}, $a_1$ exists when the experimental propensity score $\E_{\exp}(D|X)$ is bounded away from zero and one almost surely, and when the covariate density ratio $\frac{\density(X)}{\density_{\exp}(X)}$ is bounded above almost surely. The former condition is standard for treatment effects, while the latter condition is standard for covariate shifts. These are familiar conditions for regular estimation in cross sectional causal inference.

For recursively identified functionals, $T>1$ and Assumption~\ref{assumption:existence} imposes stronger regularity. Concretely, in Example~\ref{ex:surrogate}, existence of $a_2$ is guaranteed when the density ratio $\frac{\density_{\exp}(X,S)}{\density_{\obs}(X,S)}$ is bounded above almost surely. In other examples, it is guaranteed when propensity scores conditional on various trajectories are bounded away from zero and one. While overlap is far from a weak condition, it is standard in the literature, and in this sense we demonstrate how existing assumptions imply automatic and recursive orthogonality.

\begin{theorem}[Recursive orthogonality]\label{theorem:orthogonality}
    Consider the general notation of Definition~\ref{def:linear}. Suppose Assumption~\ref{assumption:existence} holds. Then the recursively identified functional $\theta_0$ has a Neyman orthogonal estimating equation given by $ \theta_0=\mathcal{M}(f_1,...,f_T,a_1,...,a_T)$ where
    $$
   \mathcal{M}(f_1,...,f_T,a_1,...,a_T)=\underbrace{\E_0\{m_1(V,f_1)\}}_{\text{identification}}+\sum_{t=1}^T \underbrace{\E_t[a_t(H_t)\{m_{t+1}(V,f_{t+1})-f_t(H_t)\}]}_{\text{recursive debiasing}}.
    $$
    As notational conventions, $a_0=1$ and $m_{T+1}(V,f_{T+1})=Y$. The functions are recursively defined: 
    $$
    f_t(H_t)=\E_t\{m_{t+1}(V,f_{t+1})|H_t\},\quad \E_t\{a_t(H_t)\tilde{f}_t(H_t)\}=\E_{t-1}\{a_{t-1}(H_{t-1})m_t(V,\tilde{f}_t)\}.
    $$
\end{theorem}

Below, we verify that this moment condition is not only Neyman orthogonal but also doubly robust. In particular, it has the mixed bias property, so we achieve $n^{-1/2}$ inference on $\theta_0$ under product rate conditions on $(f_t,a_t)$ by appealing to known results \citep{chernozhukov2018double}.

For now, we emphasize the intuition behind our main result. The final regression is $f_T(H_T)=\E_T(Y|H_T)$, i.e. the expected outcome conditional upon the history $H_T$, using a time-specific measure. From $f_T$, the earlier regression $f_{T-1}$ is recursively defined, and so on until $f_1$. The identifying formula is a linear functional of $f_1$. For each recursive regression, we construct a debiasing term that consists of a product between a recursively defined Riesz representer and a kind of regression residual, using an appropriate measure. While the regressions recursively expect the future, in that $f_t$ is defined from the later $f_{t+1}$, the Riesz representers recursively represent the past, in that $a_t$ is constructed from the earlier $a_{t-1}$. 

Instantiating our main result for specific examples yields what appears to be a new influence function  (Example~\ref{ex:diff}), and recovers many known influence functions which have been analytically derived in previous work on a case-by-case basis. Our contribution is a general construction, that is automatic in nature, in terms of recursive Riesz representers.

\subsection{Automatic and recursive inference}

First, we verify that the estimating equation in Theorem~\ref{theorem:orthogonality} satisfies the mixed bias property \citep{rotnitzky2021characterization}, implying that it is doubly robust.

\begin{lemma}[Mixed bias]\label{lemma:bias}
    Consider placeholder functions $(\tilde{f}_t,\tilde{a}_t)$ with the same arguments as the true functions $(f_t,a_t)$. Then the bias of using the placeholders instead of the true functions is
$$\mathcal{M}(\tilde{f}_1,\ldots,\tilde{f}_T,\tilde{a}_1,\ldots,\tilde{a}_T)-\theta_0
=\sum_{t=1}^T \E_t[\{\tilde{a}_t(H_t)-a_t(H_t)\}\{m_{t+1}(V,\tilde{f}_{t+1}-f_{t+1})-\tilde{f}_t(H_t)+f_t(H_t)\}]
  $$
     with the convention $m_{T+1}(V,\tilde{f}_{T+1}-f_{T+1})=0$.
\end{lemma}

We see that the bias vanishes completely in several settings. First, it vanishes when all the regressions are correctly specified, i.e.  $\tilde{f}_t=f_t$ for all $t=1,...,T$. Second, it vanishes when all the Riesz representers are correctly specified, i.e. $\tilde{a}_t=a_t$ for all $t=1,...,T$. Third, it vanishes when, for each time period $t$, either (i) $\tilde{a}_t=a_t$ or (ii) $\tilde{f}_{t+1}=f_{t+1}$ and $\tilde{f}_t=f_t$. In this sense, it has a rich double robustness property.

More generally, the bias has a product structure, which is familiar from cross sectional models. Unlike cross sectional models, it is a sum of products across time periods, and each term in the sum is with respect to a time-specific measure.

The main consequence of a Neyman orthogonal estimating equation is that it leads to regular estimation and inference on the functional of interest $\theta_0$ while tolerating machine learning for the functions $(f_t,a_t)$. We analyze an algorithm that combines our estimating equation with  sample splitting. Suppose we begin with $2n$ independent and identically distributed observations.

\begin{algorithm}[Recursively debiased machine learning]\label{algo:dml}
    Randomly the sample into equally sized $\train$ and $\test$ folds, each of size $n$.
    \begin{enumerate}
        \item For each time period $t=1,...,T$, estimate $\hat{f}_t$ and $\hat{a}_t$ using observations in $\train$.
        \item For each time period $t=0,...,T$ and each observation $i\in\test$, do the following.
        \begin{enumerate}
            \item  For $t=0$, take $\hat{\phi}_{0,i}=m_1(V_i,\hat{f}_1)$. For $t=1,\dots,T-1$, take  $\hat{\phi}_{t,i}=\hat{a}_t(H_{t,i})\{m_{t+1}(V_i,\hat{f}_{t+1})-\hat{f}_t(H_{t,i})\}$, using the convention $m_{T+1}(V_i,\hat{f}_{T+1})=Y_i$.
            \item Let $R_{t,i}$ indicate whether unit $i$ belongs to the subgroup for the measure $\E_t(\cdot)$.
        \end{enumerate}
        \item Estimate the functional as
        $
       \hat{\theta}=\sum_{t=0}^T \hat{\theta}_t,$ where $\hat{\theta}_t=\frac{1}{n_t}\sum_{i\in\test}R_{t,i} \hat{\phi}_{t,i}$ and $n_t=\sum_{i\in\test} R_{t,i}.
        $
        \item Estimate the asymptotic variance as
        $
        \hat{\sigma}^2=\frac{1}{n}\sum_{i\in \test} \hat{\psi}_i^2,$ where $\hat{\psi}_i= \sum_{t=0}^T \frac{n}{n_t} R_{t,i}(\hat{\phi}_{t,i}-\hat{\theta}_t).
        $
        \item Return the $95\%$ confidence interval 
        $
        \hat{\theta}\pm 1.96 n^{-1/2} \hat{\sigma}.
        $
    \end{enumerate}
\end{algorithm}

Algorithm~\ref{algo:dml} is a modest extension of debiased machine learning to accommodate changes of measure across time. For ease of exposition, we focus on sample splitting, but all of our theoretical results go through with cross fitting, for any fixed number of splits in the sample. 

In the remainder of this section, we articulate three assumptions under which Algorithm~\ref{algo:dml} is guaranteed to perform well: continuity, product rates, and bounded variances. To lighten notation, define the mean square norm with respect to the time $t$ measure: $\|V\|_t=\{\E_t(V^2)\}^{1/2}$.

\begin{assumption}[Continuity]\label{assumption:continuity}
    There exists a constant $\bar{M}<\infty$ such that, for each time period $t=0,...,T-1$, $\|m_{t+1}(\cdot,\tilde{f}_{t+1})\|_t\leq \bar{M} \| \tilde{f}_{t+1}\|_{t+1} $ for all placeholders $\tilde{f}_{t+1}$ with finite variance.
\end{assumption}

Assumption~\ref{assumption:continuity} ensures that if two time-specific regression functions are close, then their time-specific counterfactuals are close. The role of Assumption~\ref{assumption:continuity} is clear from Lemma~\ref{lemma:bias}: it allows us to control $\|m_{t+1}(\cdot,\hat{f}_{t+1}-f_{t+1})\|_t$ by appealing to rate conditions of the form $\|\hat{f}_{t+1}-f_{t+1}\|_{t+1}$. Similar to Assumption~\ref{assumption:existence}, Assumption~\ref{assumption:continuity} holds under familiar overlap conditions. Concretely, in Example~\ref{ex:surrogate}, we need $\|\tilde{f}_1(\cdot,1)-\tilde{f}_1(\cdot,0)\|\leq \bar{M} \|\tilde{f}_1\|_{\exp}$ and $\|\tilde{f}_2\|_{\exp}\leq \bar{M} \|\tilde{f}_2\|_{\obs}$. These conditions are satisfied under the overlap conditions discussed below Assumption~\ref{assumption:existence}, namely a propensity score bounded away from zero and one, and density ratios bounded above.

\begin{assumption}[Product rates]\label{assumption:rates}
    For each time period $t=1,...,T$,  the function estimators individually satisfy $\|\hat{f}_t-f_t\|_t=o_p(1)$ and $\|\hat{a}_t-a_t\|_t=o_p(1)$. Moreover, they jointly satisfy
$$
n_t^{1/2} \|\hat{a}_t-a_t\|_t \|\hat{f}_t-f_t\|_t=o_p(1),\quad n_t^{1/2} \|\hat{a}_t-a_t\|_t \|\hat{f}_{t+1}-f_{t+1}\|_{t+1}=o_p(1).
$$
The former holds for $t=1,...,T$ while the latter holds for $t=1,...,T-1$.
\end{assumption}

Assumption~\ref{assumption:rates} imposes consistency of the function estimators, as well as two joint rate conditions. The joint rates have product structure, allowing a trade-off: $\|\hat{a}_t-a_t\|_t$ can converge more slowly, as long as $ \|\hat{f}_t-f_t\|_t$ and $\|\hat{f}_{t+1}-f_{t+1}\|_{t+1}$ converge more quickly, and vice versa. The necessity of two joint rate conditions rather than one is immediate from Lemma~\ref{lemma:bias}, Assumption~\ref{assumption:continuity}, and the Cauchy-Schwarz inequality. Explicitly deriving the rates which satisfy Assumption~\ref{assumption:rates} is the focus of Section~\ref{sec:riesz} below. 

Here, we impose consistency of $\hat{a}_t$ and $\hat{f}_t$ to emphasize inference at the rate $n^{-1/2}$. Because Lemma~\ref{lemma:bias} proves double robustness, the recursive functional estimator $\hat{\theta}$ in Algorithm~\ref{algo:dml} will continue to be consistent when a subset of function estimators are mis-specified, though typically at a rate slower than $n^{-1/2}$. We set aside those details for brevity.

\begin{assumption}[Bounded variances]\label{assumption:variance}
We place three regularity conditions.
\begin{enumerate}
    \item  There exists a constant $\bar{\sigma}^2$ such that, for each time period $t=1,...,T$, $\E_t [\{m_{t+1}(V,f_{t+1})-f_t(H_t)\}^2|H_t]\leq \bar{\sigma}^2$ almost surely.
        \item  There exists a constant $\bar{a}$ such that, for each $t=1,...,T$, $|a_t(H_t)|\leq \bar{a}$ almost surely.
    \item Moreover, for each time period $t=1,...,T$, $|\hat{a}_t(H_t)|\leq \bar{a}$ almost surely.
\end{enumerate}
\end{assumption}

Assumptions~\ref{assumption:variance}.1 and Assumption~\ref{assumption:variance}.2 are weak regularity conditions we use to prove asymptotic normality. Assumption~\ref{assumption:variance}.1 bounds the variance of each regression residual. Assumption~\ref{assumption:variance}.2 bounds the Riesz representer in absolute value, which again holds under familiar overlap conditions. Concretely, in Example~\ref{ex:surrogate}, $\bar{a}$ exists when a propensity score is bounded away from zero and when density ratios are bounded above. 

Assumption~\ref{assumption:variance}.3 is unnecessary for asymptotic normality. It simplifies the argument for consistency of the estimator $\hat{\sigma}^2$ for the asymptotic variance $\sigma^2$. It can be eliminated by modifying $\hat{\sigma}^2$ in Algorithm~\ref{algo:dml}: instead of using $\hat{a}_t(H_t)$ directly, trim $|\hat{a}_t(H_t)|$ to be at most $\bar{a}'$, where $\bar{a}'$ is a slowly diverging sequence. We set aside these details for brevity.

\begin{corollary}[Recursive inference]\label{cor:dml}
    Suppose Assumptions~\ref{assumption:existence},~\ref{assumption:continuity},~\ref{assumption:rates}, and~\ref{assumption:variance} hold. Suppose that each $n_t$ is of order $n$. Then the estimator in Algorithm~\ref{algo:dml} is consistent and asymptotically normal, with a valid confidence interval: 
    $$
    \hat{\theta}=\theta_0+o_p(1),\quad n^{1/2}(\hat{\theta}-\theta_0)\rightsquigarrow \mathcal{N}(0,\sigma^2),\quad \P(\theta_0\in [\hat{\theta}\pm 1.96 n^{-1/2}\hat{\sigma}])\rightarrow 0.95,
    $$
    where $\sigma^2$ is the population analogue of $\hat{\sigma}^2$ defined in Algorithm~\ref{algo:dml}, and assumed to be nonzero.
\end{corollary}

Our core contribution is the automatic and recursive construction of the Neyman orthogonal estimating equation in Theorem~\ref{theorem:orthogonality}; after deriving it, Corollary~\ref{cor:dml} is an extension of known debiased machine learning theory \citep{chernozhukov2018double,chernozhukov2023simple}. This extension clarifies the main practical consequence of our orthogonalizing construction: inference at the rate $n^{-1/2}$. See \cite{li2023efficient} for a discussion of semiparametric effiency.

To ease exposition, we impose that the effective sample size $n_t$ corresponding to each measure $\E_t(\cdot)$ is of the same order. This condition amounts to proportional selection: $\E(R_t)$ is bounded away from zero, so each $n_t$ is proportional to $n$. This simplification ensures that the final rate is $n^{-1/2}$. Our analysis naturally extends to the case where $n_t$ are not proportional. In such case, the rate of inference is $(\min_t n_t)^{-1/2}$ rather than $n^{-1/2}$ \citep{chernozhukov2023automatic}.
\section{Automatic and recursive Riesz representer}\label{sec:riesz}

For inference, Assumption~\ref{assumption:rates} requires control of $\|\hat{f}_t-f_t\|_t$ and $\|\hat{a}_t-a_t\|_t$. Since each $f_t$ is a recursively defined regression, a natural estimator $\hat{f}_t$ is obtained by recursively applying an out-of-the-box machine learning regression estimator. The more interesting object is $a_t$, a new and recursively defined Riesz representer. We propose a novel estimator $\hat{a}_t$ by recursing an existing estimator for the Riesz representer in cross-sectional causal inference. Our secondary contribution is to prove nonasymptotic rates of convergence for both $\|\hat{a}_t-a_t\|_t$ and  $\|\hat{f}_t-f_t\|_t$, allowing for general machine learning.

\subsection{Intuition: Long term treatment effect} 
\label{subsec:intuition-long-term-treatment-effect}
For intuition, we return to Example~\ref{ex:surrogate}. We previously characterized the Riesz representers as
\begin{align*}
    a_1(X,D)&=\frac{\density(X)}{\density_{\exp}(X)}\left\{\frac{D}{\E_{\exp}(D|X)}-\frac{1-D}{1-\E_{\exp}(D|X)}\right\}\\
    a_2(X,S)&=\frac{\density_{\exp}(X,S)}{\density_{\obs}(X,S)}\E_{\exp}\{a_1(X,D)|X,S\}.
\end{align*}
The functional forms are complex. The ``manual'' method of estimating and inverting the components $\E_{\exp}(D|X)^{-1}$, $\density_{\exp}(X)^{-1}$, and $\density_{\obs}(X,S)^{-1}$ is computationally demanding and numerically unstable. Therefore, we propose an ``automatic'' method for directly estimating the entire functions $a_1$ and $a_2$ by appealing to the Riesz representation theorem.

Recall that $a_1$ is the base-case Riesz representer in the sense that, for all placeholders $\tilde{f}_1$,
$$
\E_{\exp}\{a_1(X,D)\tilde{f}_1(X,D)\}=\E\{\tilde{f}_1(X,1)-\tilde{f}_1(X,0)\}.
$$
Trivially, $a_1\in \argmin_{\tilde{a}_1} \|\tilde{a}_1-a_1\|_{\exp}^2$. Expanding the square and using this equation,
\begin{align*}
    a_1&\in \argmin_{\tilde{a}_1} \left[\|\tilde{a}_1\|^2_{\exp}-2\E_{\exp}\{\tilde{a}_1(X,D)a_1(X,D)\}+\|a_1\|^2_{\exp}\right]\\
    &=\argmin_{\tilde{a}_1} \left[\E_{\exp}\{\tilde{a}_1(X,D)^2\}-2 \E\{\tilde{a}_1(X,1)-\tilde{a}_1(X,0)\}\right].
\end{align*}
The empirical analogue is a feasible and known loss for estimation \citep{chernozhukov2023automatic}.

Recall that $a_2$ is the recursive Riesz representer in the sense that, for all placeholders $\tilde{f}_2$,
$$
\E_{\obs}\{a_2(X,S)\tilde{f}_2(X,S)\}=\E_{\exp}\{a_1(X,D)\tilde{f}_2(X,S)\}.
$$
Now $a_2\in \argmin_{\tilde{a}_2} \|\tilde{a}_2-a_2\|_{\obs}^2$. As before, expanding the square and using this equation, 
\begin{align*}
    a_2&\in \argmin_{\tilde{a}_2} \left[\|\tilde{a}_2\|^2_{\obs}-2\E_{\obs}\{\tilde{a}_2(X,S)a_2(X,S)\}+\|a_2\|^2_{\obs}\right]\\
    &=\argmin_{\tilde{a}_2} \left[\E_{\obs}\{\tilde{a}_2(X,S)^2\}-2 \E_{\exp}\{a_1(X,D)\tilde{a}_2(X,S)\}\right].
\end{align*}
However, the empirical analogue is no longer feasible: it contains the unknown function $a_1$. 

In this work, we propose the recursive estimator
$$
\hat{a}_2\in \argmin_{\tilde{a}_2} \left[\hat{\E}_{\obs}\{\tilde{a}_2(X,S)^2\}-2 \hat{\E}_{\exp}\{\hat{a}_1(X,D)\tilde{a}_2(X,S)\}\right].
$$
Crucially, $\hat{a}_1$ appears in the loss for $\hat{a}_2$. Therefore, the estimation error compounds across time periods, which is a new technical challenge. Previous mean square analyses of Riesz estimation error apply to $\hat{a}_1$ but not $\hat{a}_2$; new theory is required.

\subsection{Estimation with generic machine learning}

In general, for the setting of Definition~\ref{def:linear}, we estimate $a_t$ from $a_{t-1}$. By Assumption~\ref{assumption:existence}, we have
 $$
    \E_t\{a_t(H_t)\tilde{f}_t(H_t)\}=\E_{t-1}\{a_{t-1}(H_{t-1})m_t(V,\tilde{f}_t)\}.
    $$
    Since $a_t\in\argmin_{\tilde{a}_t}\|\tilde{a}_t-a_t\|^2_t$, our technique generalizes to 
    \begin{align*}
        a_t&\in \argmin_{\tilde{a}_t} \left[ \|\tilde{a}_t\|^2_t-2\E_t\{\tilde{a}_t(H_t)a_t(H_t)\}+\|a_t\|^2_t\right] \\
        &=\argmin_{\tilde{a}_t}  \left[\E_t\{\tilde{a}_t(H_t)^2\}-2\E_{t-1}\{a_{t-1}(H_{t-1})m_t(V,\tilde{a}_t)\}\right].
    \end{align*}
We define our estimator accordingly.

\begin{algorithm}[Recursive Riesz representer]\label{algo:riesz}
    Use the $\train$ fold of size $n$. Fix a function space for estimation $\mathcal{A}$. For each time period $t=1,...,T$, recursively estimate
    $$
    \hat{a}_t\in\argmin_{\tilde{a}_t\in\mathcal{A}}\left[\hat{\E}_t\{\tilde{a}_t(H_t)^2\}-2\hat{\E}_{t-1}\{\hat{a}_{t-1}(H_{t-1})m_t(V,\tilde{a}_t)\}\right],\quad \hat{a}_0=1.
    $$
    Here, $\hat{\E}_t(\cdot)$ averages over the observations in $\train$ which correspond to the subgroup in $\E_t(\cdot)$.
\end{algorithm}

Algorithm~\ref{algo:riesz} is a recursive generalization of the Riesz regression \citep{chernozhukov2022automatic,chernozhukov2021automatic}, allowing for changes of measure \citep{chernozhukov2023automatic}. This generalization is necessary to handle the recursive functionals in Definition~\ref{def:linear}; earlier works on Riesz regression only apply to cross-sectional models. For simplicity, we do not have a penalty term, however that can be introduced with a longer proof \citep{chernozhukov2021automatic}.

Relative to some earlier works on balancing weights \citep{kallus2021optimal,viviano2026dynamic}, we make four contributions. First, we handle all recursive functionals satisfying Definition~\ref{def:linear}, including the many examples in Section~\ref{sec:examples}, rather than only Example~\ref{ex:time}. Second, we allow the estimation space $\mathcal{A}$ to be generic, e.g. neural nets and random forests, rather than only the lasso or reproducing kernel Hilbert space. Third, we do not restrict the causal model beyond the familiar assumption of overlap. By contrast, \citet[eq.3]{kallus2021optimal} impose a parametric family of expected potential outcomes, and \citet[Assumption 4.1(C)]{viviano2026dynamic} impose linearity of expected potential outcomes. Fourth, we derive mean square convergence rates, to which we now turn. 

Our analysis is agnostic about the spaces in which $f_t$ and $\hat{f}_t$ reside. 
If, instead, the researcher knew that $f_t$ and $\hat{f}_t$ belonged to the same closed linear subspace, then it would be efficient in this restricted model to choose $\mathcal{A}$ as the same subspace \citep{chernozhukov2022debiased,li2023efficient}. 
For example, if $f_t$ is known to not depend on a component of $H_t$, then that component should be excluded from estimation of $\hat{f}_t$ and $\hat{a}_t$.

\subsection{Sharp rate analysis}

The main assumption for our rate analysis is that the function space $\mathcal{A}$ used for estimation is not too complex. Formally, it has a complexity measure called a critical radius, which will appear in the rate. Appendix~\ref{sec:riesz_proof} provides background on critical radius theory.

\begin{assumption}[Critical radii]\label{assumption:critical}
For each time period $t=1,...,T$, suppose $a_t\in\mathcal{A}$. As a convention, take $\tilde{a}_0=a_0=1$.  Define the function spaces
\begin{align*}
    \mathcal{F}_t&=[h_t\mapsto \kappa\{\tilde{a}_t(h_t)-a_t(h_t)\}:\quad \tilde{a}_t\in\mathcal{A},\quad \kappa\in[0,1] ] \\
      \mathcal{F}'_{t}&=[v\mapsto
\kappa
\tilde{a}_{t-1}(h_{t-1})
m_t(v,\tilde{a}_t-a_t):\quad 
\tilde{a}_t,\tilde{a}_{t-1}\in\mathcal{A},\quad
\kappa\in[0,1]].
\end{align*}
Assume these spaces and $\mathcal{A}$ contain functions uniformly bounded by $\bar{a}$. Suppose that $\mathcal{F}_t$ has critical radius $\delta_t$ when evaluated under the measure of $\E_t(\cdot)$, and that $\mathcal{F}_t'$ has critical radius $\delta'_t$ when evaluated under the measure of $\E_{t-1}(\cdot)$. Finally, suppose $\delta_t^2\geq c \frac{\ln\{\ln(n_t)\}}{n_t}$ for some constant $c>0$.
\end{assumption}

Critical radii are known for a variety of machine learning function spaces, such as simple neural networks and random forests. If $\mathcal{A}$ is parametric, then $\delta_t=O(n_t^{-1/2})$. For high dimensional linear functions with $s$ nonzero coefficients in $p$ dimensions, $\delta_t=O[\ln(n_t)\{s\ln(p)/n_t\}^{1/2}]$. For a neural network with rectified linear activation, depth $L$, and $W$ parameters, $\delta_t=\{LW \ln(W)\ln(n_t)\}n_t^{-1/2}$. See e.g. \cite{foster2023orthogonal} for a recent review.

The critical radius condition may be viewed as a relaxation of the Donsker condition used by \cite{ai2003efficient}, among others.  See \citet[eq. 1 and 6]{chernozhukov2026adversarial} for a detailed comparison between the critical radius condition and the Donsker condition in terms of entropy integrals, building on \citet[Corollary 14.3]{wainwright2019high}.

Assumption~\ref{assumption:critical} makes some simplifications for readability, which can be relaxed. First, it assumes correct specification: $a_t\in\mathcal{A}$. Our results can extend to accommodate approximation error. Second, it uses the same space $\mathcal{A}$ across time periods. This choice is only to lighten notation. Third, we impose uniform boundedness of functions, which future work may relax to bounded moments. The lower bound on $\delta_t^2$ is an non-binding artifact of our proof technique.\footnote{The condition arises when using Bernstein-style arguments in nonlinear settings \citep{wainwright2019high}.}

\begin{theorem}[Recursive estimation error]\label{theorem:rate}
    Suppose Assumptions~\ref{assumption:existence},~\ref{assumption:continuity},~\ref{assumption:variance}.2, and~\ref{assumption:critical} hold.\footnote{To simplify the constants, we also assume $\bar{a},\bar{M}\geq 1$, but this can be relaxed.} Then with probability $1-\eta$, for every $t=1,...,T$ and some universal constant $C<\infty$,
    $$
    \|\hat{a}_t-a_t\|^2_{t}\leq C \cdot (\bar{a}\bar{M})^2 \cdot \left\{  \max(\delta_t,\delta_t')^2+\frac{\ln(T/\eta)}{\min(n_t,n_{t-1})}+\|\hat{a}_{t-1}-a_{t-1}\|^2_{t-1}\right\}.
    $$
    Consequently, for a fixed number of time periods $T$, recursing backwards in time gives
    $$
    \|\hat{a}_t-a_t\|_t^2\leq T\cdot \{C(\bar{a}\bar{M})^{2}\}^T \cdot \left[\max_{s\in\{1,...,T\}} \left\{  \max(\delta_s,\delta_s')^2\right\}+\frac{\ln(T/\eta)}{\min_{s\in\{0,...,T\}}(n_s)}\right].
    $$
\end{theorem}

Theorem~\ref{theorem:rate} is our secondary contribution: a non-asymptotic bound on the mean square error of the recursive Riesz representer. The initial display is highly interpretable. The pre-factors reflect the strength of overlap. The first term reflects the complexity of the estimator. The second term reflects the effective sample size for the time period.  The third term reflects the estimation error inherited from the previous time period. 

The final term is not present in previous rate analyses of Riesz representers, which do not have a recursive structure. Our key insight is that estimation error from the past compounds in an additive way. Therefore, when we recurse backwards in time, we can simply take a maximum over time periods, given by the latter display. Taking a maximum over a finite number of time periods is allowed. Future work may study a growing number of time periods.

While Theorem~\ref{theorem:rate} is stated as a rate for the recursive Riesz representer $\|\hat{a}_t-a_t\|^2_t$, it also implies a rate for the recursive regression $\|\hat{f}_t-f_t\|^2_t$. In particular, the regression function may be viewed as a special case of a Riesz representer. For intuition, consider the cross sectional setting: $f(H)=\E(Y|H)$ is the Riesz representer of the functional $\tilde{f}\mapsto \E\{Y\tilde{f}(H)\}$.\footnote{In more detail, the recursive regression estimator $\hat{f}_t\in \argmin_{\tilde{f}_t\in\mathcal{F}}\hat{\E}_t[\{m_{t+1}(V,\hat{f}_{t+1})-\tilde{f}_t(H_t)\}^2]$ can be expressed as $\hat{f}_t\in \argmin_{\tilde{f}_t\in\mathcal{F}}[\hat{\E}_t\{\tilde{f}_t(H_t)^2\}-2\hat{\E}_t \{m_{t+1}(V,\hat{f}_{t+1})\tilde{f}_t(H_t)\}]$, matching the structure of Algorithm~\ref{algo:riesz}.}

Therefore, Theorem~\ref{theorem:rate} provides the rates required by Assumption~\ref{assumption:rates}. This result is necessary to operationalize the Neyman orthogonal estimating equation in Theorem~\ref{theorem:orthogonality}. Together, Theorems~\ref{theorem:orthogonality} and~\ref{theorem:rate} provide an end-to-end framework for automatic estimation and inference of recursive functionals. 
\section{Simulation evidence and real-world applications}\label{sec:empirics}

We apply our method for automatic and recursive inference to synthetic and real data. 
With synthetic data, we demonstrate that our method reduces bias and improves precision relative to some previous methods that have manual debiasing or balancing weights in limited function spaces. 
With real data, our method recovers a long-term experimental benchmark when combining short term experimental data with long-term observational data on job training.
With real data, our method provides flexible yet precise estimates in a dynamic difference-in-difference analysis of the minimum wage. Overall, our method is highly practical in two prototypical settings for labor economic research.

\subsection{Bias reductions and precision gains}

Because the majority of methodological work on recursive functionals focuses on the time-varying treatment effect (Example~\ref{ex:time}), so does our first exercise. Adapting simulation designs from \cite{bradic2024high,chernozhukov2026adversarial}, we compare three types of methods: 
(i) oracle debiased machine learning with known regressions and propensity scores; 
(ii) manual debiased machine learning with estimated regressions and propensity scores; (iii) our method: automatic debiased machine learning with estimated regressions and  recursive Riesz representers. The lasso version of our method is similar to a previous balancing weight method \citep{viviano2026dynamic}, though derived without assuming linear potential outcomes. Unlike earlier work, we further allow random forest and neural network versions, and handle additional functionals. See Appendix~\ref{sec:sim} for implementation details and similar results for several other variations of manual debiased machine learning.

Tables~\ref{tab:time_main_linear} and~\ref{tab:time_main_nonlinear} summarize our findings for linear and nonlinear simulation designs, respectively. 
For each sample size and method, we report several metrics across $500$ simulations: the bias and root mean square error of the point estimates, as well as the coverage and length of the confidence intervals.
The oracle is infeasible, because regressions and propensity scores are unknown to the economist, but it serves as a benchmark for evaluating the feasible methods.

\begin{table}
\centering
\caption{Linear time-varying treatment effect simulation}
\label{tab:time_main_linear}
\small
\setstretch{1.0}
\setlength{\tabcolsep}{15pt}
\begin{tabular}{clcccc}
\toprule
\toprule
$n$ & Method & Bias & RMSE & Coverage & CI length \\
\midrule
\multirow{5}{*}{500}
 & Oracle       &  0.023 & 0.245 & 0.932 & 0.877 \\
 & Manual-Lasso &  0.076 & 0.611 & 0.942 & 1.295 \\
 & Auto-Lasso   &  0.023 & 0.229 & 0.928 & 0.815 \\
 & Auto-RF      &  0.019 & 0.237 & 0.858 & 0.711 \\
 & Auto-NN      & -0.238 & 0.333 & 0.744 & 0.799 \\
\cmidrule(lr){1-6}
\multirow{5}{*}{1000}
 & Oracle       &  0.000 & 0.162 & 0.942 & 0.623 \\
 & Manual-Lasso &  0.012 & 0.248 & 0.942 & 0.837 \\
 & Auto-Lasso   & -0.003 & 0.149 & 0.938 & 0.564 \\
 & Auto-RF      &  0.001 & 0.156 & 0.894 & 0.506 \\
 & Auto-NN      & -0.039 & 0.177 & 0.906 & 0.637 \\
\cmidrule(lr){1-6}
\multirow{5}{*}{2000}
 & Oracle       &  0.005 & 0.110 & 0.970 & 0.441 \\
 & Manual-Lasso & -0.003 & 0.238 & 0.976 & 0.650 \\
 & Auto-Lasso   &  0.008 & 0.100 & 0.958 & 0.397 \\
 & Auto-RF      &  0.005 & 0.135 & 0.906 & 0.368 \\
 & Auto-NN      &  0.002 & 0.148 & 0.962 & 0.564 \\
\bottomrule
\bottomrule
\end{tabular}
\end{table}

Table~\ref{tab:time_main_linear} shows that, even when manual debiased machine learning is correctly specified, the automatic approach is more precise in finite samples. 
Relative to manual lasso, the automatic lasso reduces root mean square error by 40--63\% and confidence interval length by 33--39\% while maintaining similar coverage.  
Even when both methods use lasso as the function space, the automatic approach improves precision. Table~\ref{tab:time_appendix_linear} presents several additional estimators.

\begin{table}
\centering
\caption{Nonlinear time-varying treatment effect simulation}
\label{tab:time_main_nonlinear}
\small
\setstretch{1.0}
\setlength{\tabcolsep}{15pt}
\begin{tabular}{clcccc}
\toprule
\toprule
$n$ & Method & Bias & RMSE & Coverage & CI length \\
\midrule
\multirow{5}{*}{500}
 & Oracle       & -0.013 & 0.268 & 0.956 & 1.049 \\
 & Manual-Lasso &  0.143 & 0.298 & 0.902 & 1.048 \\
 & Auto-Lasso   &  0.171 & 0.311 & 0.910 & 1.027 \\
 & Auto-RF      &  0.080 & 0.285 & 0.930 & 1.051 \\
 & Auto-NN      &  0.234 & 0.356 & 0.822 & 0.999 \\
\cmidrule(lr){1-6}
\multirow{5}{*}{1000}
 & Oracle       & -0.003 & 0.197 & 0.934 & 0.739 \\
 & Manual-Lasso &  0.150 & 0.248 & 0.882 & 0.735 \\
 & Auto-Lasso   &  0.180 & 0.264 & 0.840 & 0.721 \\
 & Auto-RF      &  0.063 & 0.214 & 0.910 & 0.753 \\
 & Auto-NN      &  0.073 & 0.217 & 0.928 & 0.762 \\
\cmidrule(lr){1-6}
\multirow{5}{*}{2000}
 & Oracle       & -0.016 & 0.132 & 0.954 & 0.524 \\
 & Manual-Lasso &  0.138 & 0.189 & 0.818 & 0.519 \\
 & Auto-Lasso   &  0.171 & 0.214 & 0.748 & 0.512 \\
 & Auto-RF      &  0.028 & 0.138 & 0.952 & 0.541 \\
 & Auto-NN      &  0.063 & 0.151 & 0.914 & 0.544 \\
\bottomrule
\bottomrule
\end{tabular}
\end{table}

Table~\ref{tab:time_main_nonlinear} shows that, in challenging nonlinear settings, the random forest and neural network versions of our method are critical improvements. Only our automatic random forest method and our automatic neural network method have vanishing bias and nearly nominal coverage in large samples. By contrast, the manual and automatic lasso methods have large, persistent biases and undercover. Even in the well-studied setting of Example~\ref{ex:time} for which previous balancing weight estimators exist, our method allows for more flexibility, which is essential for nominal coverage.  Table~\ref{tab:time_appendix_nonlinear} presents several additional estimators.

For our second exercise, we consider dynamic difference-in-differences (Example~\ref{ex:diff}), due to the rapid growth in empirical economic research using such methods. To the best of our knowledge, only parametric methods based on least squares have been previously proposed for this setting.\footnote{The nonparametric method of \cite{shahn2022structural} is for a difference-in-differences model without the recursive structure of Definition~\ref{def:linear} and Example~\ref{ex:diff}.} Therefore, we compare (i) least squares estimation and (ii) our method based on recursive Riesz representers in Table~\ref{tab:diff_main}. We report the same metrics as before, again across $500$ simulations. See Appendix~\ref{sec:sim} for implementation details.

\begin{table}
\centering
\caption{Dynamic difference-in-differences simulation}
\label{tab:diff_main}
\small
\setstretch{1.0}
\setlength{\tabcolsep}{15pt}
\begin{tabular}{clcccc}
\toprule
\toprule
$n$ & Method & Bias & RMSE & Coverage & CI length \\
\midrule
\multirow{5}{*}{500}
 & OLS         & -0.256 & 0.363 & 0.842 & 1.031 \\
 & Auto-Lasso  &  0.107 & 0.279 & 0.980 & 1.214 \\
 & Auto-RF     &  0.056 & 0.264 & 0.982 & 1.206 \\
 & Auto-NN     &  0.066 & 0.269 & 0.986 & 1.243 \\
\cmidrule(lr){1-6}
\multirow{5}{*}{1000}
 & OLS         & -0.248 & 0.306 & 0.746 & 0.730 \\
 & Auto-Lasso  &  0.113 & 0.217 & 0.962 & 0.858 \\
 & Auto-RF     &  0.064 & 0.198 & 0.976 & 0.853 \\
 & Auto-NN     &  0.071 & 0.201 & 0.972 & 0.870 \\
\cmidrule(lr){1-6}
\multirow{5}{*}{2000}
 & OLS         & -0.232 & 0.268 & 0.572 & 0.515 \\
 & Auto-Lasso  &  0.125 & 0.179 & 0.922 & 0.607 \\
 & Auto-RF     &  0.075 & 0.149 & 0.954 & 0.606 \\
 & Auto-NN     &  0.074 & 0.148 & 0.964 & 0.617 \\
\bottomrule
\bottomrule
\end{tabular}
\end{table}

Table~\ref{tab:diff_main} shows that, by including examples beyond the time-varying treatment effect, our framework is useful for empirical economics. Least squares is highly biased, with undercoverage that worsens as the sample size increases. 
By contrast, versions of our method with lasso, random forest, and neural network function spaces have small bias and nearly nominal coverage. Our framework, which includes all of the examples in Section~\ref{sec:examples},  has meaningful generality: it provides new estimators for some leading economic settings where nonparametric methods have not been previously developed.

\subsection{Effect of job training using surrogate outcomes}

As our first real-world application, we study the effect of Greater Avenues to Independence (GAIN) job training on long term earnings via surrogate analysis (Example~\ref{ex:surrogate}). We replicate the exercise of \cite{athey2025surrogate}, who ask: can an estimator that combines short term experimental data with long term observational data recover the ``oracle'' estimate obtained with long term experimental data? We compare three types of  methods: (i) the difference-in-means using ``oracle'' long term experimental data; (ii) manual debiased machine learning with explicit propensity scores; (iii) automatic debiased machine learning with recursive Riesz representers. The manual approach is from previous work, while the automatic approach is new. See Appendix~\ref{sec:app} for implementation details.

\begin{figure}
        \centering
        \includegraphics[width=0.8\linewidth]{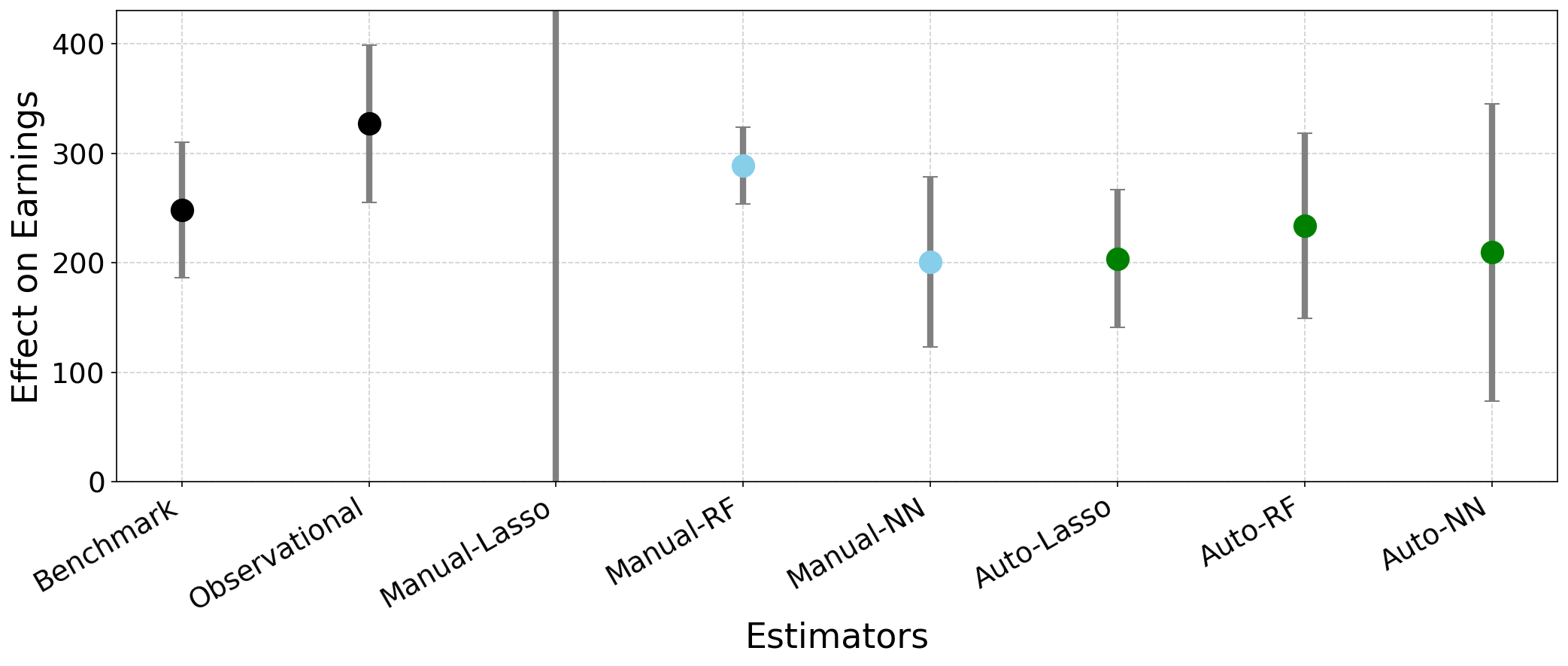}
        \label{fig:surrogate_earnings}
        \vspace{-10pt}
    \caption{Surrogate application}\label{fig:surrogate_main}
    \vspace{-10pt}
\end{figure}

Figure~\ref{fig:surrogate_main} summarizes our findings. The  benchmark estimate using long term experimental data is 248 dollars, with a standard error of 32 dollars. A researcher using only long term observational data would overestimate the effect as 327 dollars due to selection bias. Among the various debiased machine learning estimates, our automatic random forest estimate is the closest to the experimental benchmark: it is 234 dollars, only 14 dollars away. By contrast, the manual random forest estimate is 41 dollars too high in the direction of the selection bias. Similarly, the automatic neural network estimate is closer to the benchmark than the manual neural network estimate, and the automatic lasso estimate is much closer to the benchmark than the manual lasso estimate.

\subsection{Effect of minimum wage by dynamic difference-in-differences}

\begin{figure}
    \centering
    \begin{subfigure}[b]{0.48\textwidth}
        \centering
        \includegraphics[width=\linewidth]{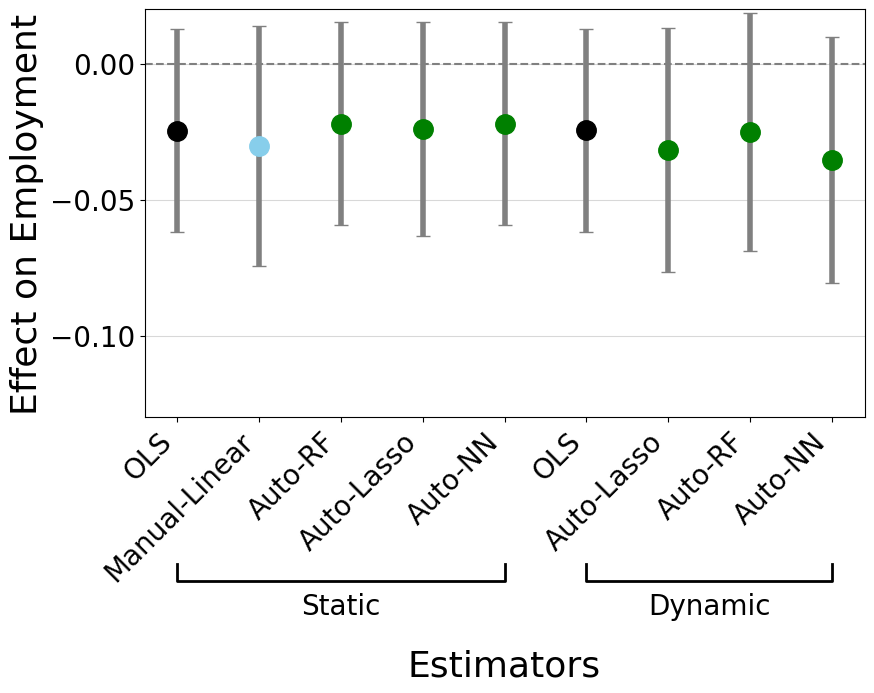
        }
        \caption{One year post-treatment}
        \label{fig:did_2004}
    \end{subfigure}
    \hfill
    \begin{subfigure}[b]{0.48\textwidth}
        \centering
        \includegraphics[width=\linewidth]{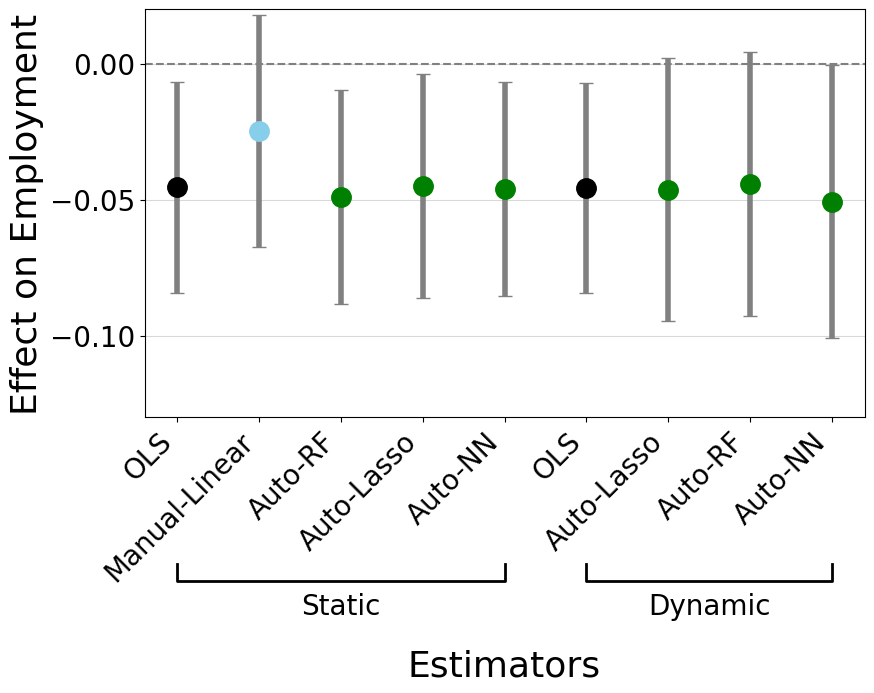
        }
        \caption{Two years post-treatment}
        \label{fig:did_2005}
    \end{subfigure}
    \begin{subfigure}[b]{0.48\textwidth}
        \centering
        \includegraphics[width=\linewidth]{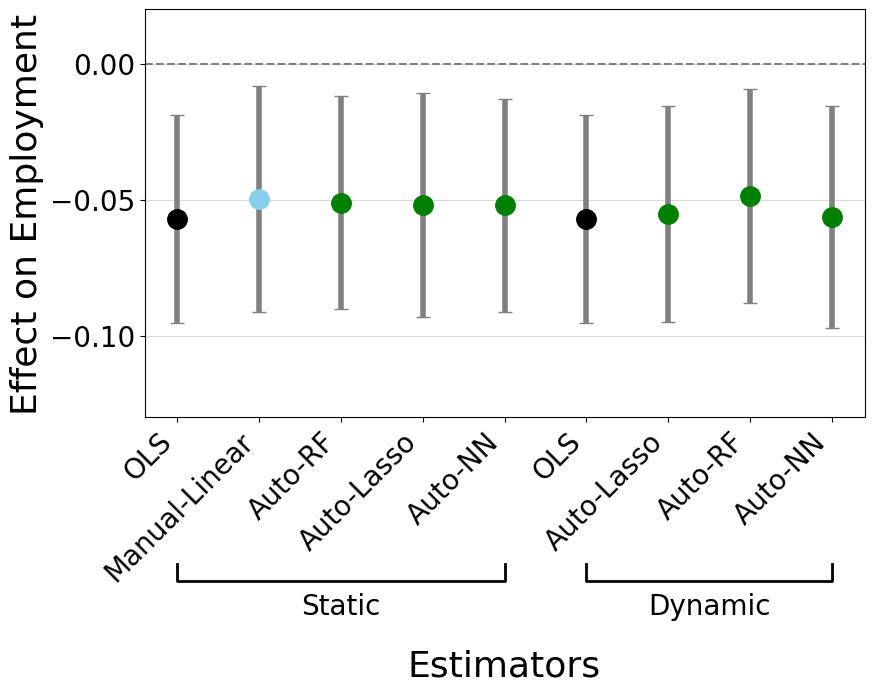}
        \caption{Three years post-treatment}
        \label{fig:did_2006}
    \end{subfigure}
    \hfill
    \begin{subfigure}[b]{0.48\textwidth}
        \centering
        \includegraphics[width=\linewidth]{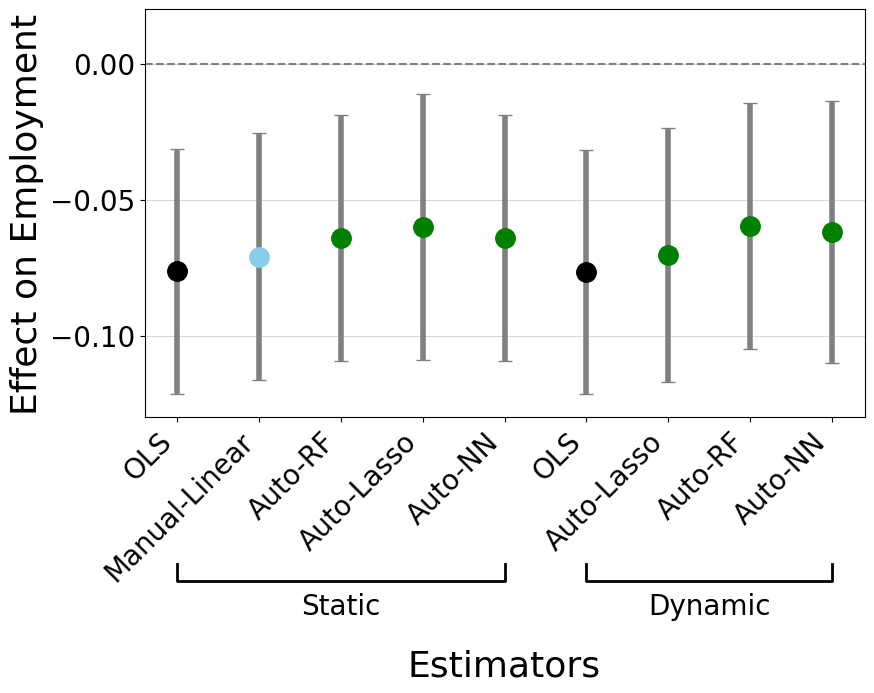
        }
        \caption{Four years post-treatment}
        \label{fig:did_2007}
    \end{subfigure}
    \caption{Dynamic difference-in-differences application}
    \label{fig:dynamic_main}
\end{figure}

As our second real-world application, we study the effect of the minimum wage on employment of teenagers via dynamic difference-in-difference analysis (Example~\ref{ex:diff}). We extend the exercise of \cite{callaway2021difference} by considering not only the standard difference-in-differences model, where parallel trends hold conditional upon pre-treatment covariates, but also the dynamic model, where parallel trends hold conditional upon pre- and post-treatment covariates. We compare two categories of methods: (i) those for static difference-in-differences, and (ii) those for dynamic  difference-in-differences. Within the static category, we implement OLS, the parametric doubly robust approach of \cite{sant2020doubly} with explicit propensity scores, and the automatic debiased machine method of \cite{chernozhukov2023automatic}. Within the dynamic category, we implement the OLS method of \cite{caetano2022difference} as well as our proposed automatic debiased machine learning method based on recursive Riesz representers. See Appendix~\ref{sec:app} for implementation details.

The primary concern of using the dynamic model rather than the standard model is that the additional richness of the causal assumptions could come at the cost of vastly inflated standard errors.  Figure~\ref{fig:dynamic_main} shows that our method allows this additional richness without inflating the standard errors. When considering the outcome to be employment one, two, three, or four years post-treatment, our proposed method for the dynamic model gives point estimates and standard errors that are almost identical to those of earlier methods for the static model. Our standard errors are only slightly larger than those of OLS, while allow rich heterogeneity. In other words, by using our method, empirical researchers may use flexible causal models with dynamic structure while preserving confidence in their findings.
\section{Discussion: Expect the future, represent the past}\label{sec:conc}

Several core models in econometrics and statistics are identified by recursively expecting the future. In this paper, we provide an automatic method for inference that recursively represents the past. Our main contribution is an automatic and recursive construction of Neyman orthogonal estimating equations. Our secondary contribution is an estimator for the recursive Riesz representers that appear in those equations.

Future work may extend our framework in several directions. While we focus on a fixed number of time periods, we conjecture that our main construction may apply to time series with a diverging number of time periods. While we focus on models with ``dynamics on observables'', we conjecture that our techniques could be applied to models with ``dynamics on unobservables'' in the form of fixed effects. More generally, there is work to do bridging the class of models in Definition~\ref{def:linear} with the rich literature on panel data. As a step in that direction, Appendix~\ref{sec:instrument} considers the extension to instrumental variables.

We conclude by adapting an aphorism attributed to L. Peter Deutsch and a haiku composed by Keisuke Hirano. 
\emph{Use representers:
to iterate is human;
to recurse, divine.}\footnote{
Deutsch's aphorism, quoted in \cite{stroustrup2013cpp}, is \emph{To iterate is human; to recurse, divine.}
Hirano's haiku, titled ``Consumption Smoothing (after L. Peter Deutsch)'',  
is \emph{Use value functions: to iterate is human; to recurse, divine.}
}

\if1\anon
{
\bibliographystyle{apalike}
\spacingset{1}{
\bibliography{bib}}
} \fi

\clearpage


\appendix
\numberwithin{figure}{section}
\numberwithin{table}{section}

\begin{center}
{\LARGE\bf Proofs}
\end{center}

Appendices~\ref{sec:orthogonality_proof} and~\ref{sec:riesz_proof} provide proofs for Sections~\ref{sec:orthogonality} and~\ref{sec:riesz}, respectively.

\section{Proofs for Section~\ref{sec:orthogonality}}\label{sec:orthogonality_proof}

\subsection{Proof of Theorem~\ref{theorem:orthogonality}}

Recall that 
$$
\mathcal{M}(\tilde{f}_1,\ldots,\tilde{f}_T,\tilde{a}_1,\ldots,\tilde{a}_T) \\
 =
\E_0\{m_1(V,\tilde{f}_1)\}
+
\sum_{t=1}^T
\E_t\left[
\tilde{a}_t(H_t)
\{m_{t+1}(V,\tilde{f}_{t+1})-\tilde{f}_t(H_t)\}
\right],
$$
with the convention $m_{T+1}(V,\tilde{f}_{T+1})=Y$. 

\begin{enumerate}
    \item We verify that this moment identifies $\theta_0$ at the true nuisance functions. By the law of iterated expectations and the recursive definition of $f_t$, for each $t=1,...,T$, 
  \begin{align*}
      \E_t\left[
a_t(H_t)
\{m_{t+1}(V,f_{t+1})-f_t(H_t)\}
\right]
&= \E_t\left[
a_t(H_t)
\E_t\{m_{t+1}(V,f_{t+1})-f_t(H_t)| H_t\}
\right] \\
&= \E_t\left[
a_t(H_t)
\{f_t(H_t)-f_t(H_t)\}
\right]
=0.
  \end{align*}
Therefore, all recursive debiasing terms have mean zero at the true nuisance functions, so
$$
\mathcal{M}(f_1,\ldots,f_T,a_1,\ldots,a_T)
=
\E_0\{m_1(V,f_1)\}
=
\theta_0.
$$
The final equality is the identifying formula in Definition~\ref{def:linear}.
    \item To verify Neyman orthogonality, we show that the pathwise derivative of $\mathcal{M}$ with respect to each nuisance function is zero at the truth.

\begin{enumerate}
    \item Fix a placeholder function $\tilde{a}_t(H_t)$ and define the path
$
a_t^{(\kappa)}
=
a_t+\kappa(\tilde{a}_t-a_t).
$
At the truth, only the $t$-th recursive debiasing term depends on $a_t$. Therefore, for $t=1,...,T$,  the pathwise derivative is 
\begin{align*}
   &\left. \frac{d}{d\kappa}
\mathcal{M}(f_1,\ldots,f_T,a_1,\ldots,a_t^{(\kappa)},\ldots,a_T) \right|_{\kappa=0}
=\E_t\left[
\{\tilde{a}_t(H_t)-a_t(H_t)\}
\{m_{t+1}(V,f_{t+1})-f_t(H_t)\}
\right] \\
&=\E_t\left[
\{\tilde{a}_t(H_t)-a_t(H_t)\}
\E_t\{m_{t+1}(V,f_{t+1})-f_t(H_t)|H_t\}
\right] \\
&= \E_t\left[
\{\tilde{a}_t(H_t)-a_t(H_t)\}
\{f_t(H_t)-f_t(H_t)\}
\right]
=0
\end{align*}
by scalar differentiation, the law of iterated expectations, and the recursive definition of $f_t$.

\item Next fix a placeholder function $\tilde{f}_t(H_t)$ and define the path
$
f_t^{(\kappa)}
=
f_t+\kappa(\tilde{f}_t-f_t).
$
At the truth, $f_t$ appears only in two places: $m_t(V,f_t)$ in the previous term, and $-f_t(H_t)$ in the current term.  Recall that $f_t\mapsto m_t(V,f_t)$ is linear by Definition~\ref{def:linear}. 

For $t=1$, the pathwise derivative is 
\begin{align*}
     \left.
\frac{d}{d\kappa}
\mathcal{M}(f_1^{(\kappa)},f_2,\ldots,f_T,a_1,\ldots,a_T)
\right|_{\kappa=0}
&=
\E_0\{m_1(V,\tilde{f}_1-f_1)\}
-
\E_1\{a_1(H_1)(\tilde{f}_1-f_1)(H_1)\}\\
&=\E_0\{m_1(V,\tilde{f}_1-f_1)\}
-
\E_0\{a_0m_1(V,\tilde{f}_1-f_1)\}\\ 
&=\E_0\{m_1(V,\tilde{f}_1-f_1)\}
-
\E_0\{m_1(V,\tilde{f}_1-f_1)\}
=0
\end{align*}
by scalar differentiation, Assumption~\ref{assumption:existence}, and the convention $a_0=1$.

For $t=2,\ldots,T$, the pathwise derivative is
\begin{align*}
    &\left.
\frac{d}{d\kappa}
\mathcal{M}(f_1,\ldots,f_t^{(\kappa)},\ldots,f_T,a_1,\ldots,a_T)
\right|_{\kappa=0} \\
&=\E_{t-1}\{a_{t-1}(H_{t-1})m_t(V,\tilde{f}_t-f_t)\}
-
\E_t\{a_t(H_t)(\tilde{f}_t-f_t)(H_t)\}
=0
\end{align*}
by scalar differentiation and Assumption~\ref{assumption:existence}. \qed
\end{enumerate}
\end{enumerate}

\subsection{Proof of Lemma~\ref{lemma:bias}}

    To lighten notation, define
$
\Delta f_t=\tilde{f}_t-f_t$ and $\Delta a_t=\tilde{a}_t-a_t.
$

By Theorem~\ref{theorem:orthogonality} and linearity of $f_t\mapsto m_t(V,f_t)$, the bias is    
\begin{align*}
&\mathcal{M}(\tilde{f}_1,\ldots,\tilde{f}_T,\tilde{a}_1,\ldots,\tilde{a}_T)
-
\mathcal{M}(f_1,\ldots,f_T,a_1,\ldots,a_T)\\
&=
\E_0\{m_1(V,\Delta f_1)\}
+
\sum_{t=1}^T
\E_t\left[
\tilde{a}_t(H_t)
\{m_{t+1}(V,\tilde{f}_{t+1})-\tilde{f}_t(H_t)\}
-
a_t(H_t)
\{m_{t+1}(V,f_{t+1})-f_t(H_t)\}
\right]\\
&=
\E_0\{m_1(V,\Delta f_1)\}
+
\sum_{t=1}^T
\E_t\left[
a_t(H_t)
\{m_{t+1}(V,\Delta f_{t+1})-\Delta f_t(H_t)\}
\right]\\
&\quad+
\sum_{t=1}^T
\E_t\left[
\Delta a_t(H_t)
\{m_{t+1}(V,f_{t+1})-f_t(H_t)\}
\right]\\
&\quad+
\sum_{t=1}^T
\E_t\left[
\Delta a_t(H_t)
\{m_{t+1}(V,\Delta f_{t+1})-\Delta f_t(H_t)\}
\right].
\end{align*}
We show that the initial terms are zero. 

\begin{enumerate}
    \item First, we show
$
\E_0\{m_1(V,\Delta f_1)\}
+
\sum_{t=1}^T
\E_t\left[
a_t(H_t)
\{m_{t+1}(V,\Delta f_{t+1})-\Delta f_t(H_t)\}
\right]
=0.
$

By Assumption~\ref{assumption:existence} and the convention $a_0=1$,
$
\E_0\{m_1(V,\Delta f_1)\}
=
\E_1\{a_1(H_1)\Delta f_1(H_1)\}.
$ 
Thus $\E_0\{m_1(V,\Delta f_1)\}$ cancels the $t=1$ term of 
$
-\E_1\{a_1(H_1)\Delta f_1(H_1)\}.
$

For $t=2,\ldots,T$, Assumption~\ref{assumption:existence} gives
$
\E_{t-1}\{a_{t-1}(H_{t-1})m_t(V,\Delta f_t)\}
=
\E_t\{a_t(H_t)\Delta f_t(H_t)\}.
$
Therefore the term
$
\E_{t-1}\{a_{t-1}(H_{t-1})m_t(V,\Delta f_t)\}
$
from the $(t-1)$-th recursive correction cancels the term
$
-\E_t\{a_t(H_t)\Delta f_t(H_t)\}
$
from the $t$-th recursive correction. These cancellations telescope across $t=1,\ldots,T$. The terminal term is zero because $m_{T+1}(V,\Delta f_{T+1})=0$.
    \item Next, we show
    $
    \sum_{t=1}^T
\E_t\left[
\Delta a_t(H_t)
\{m_{t+1}(V,f_{t+1})-f_t(H_t)\}
\right]=0.
    $

    By the law of iterated expectations and the recursive definition of $f_t$,
$$
\E_t\left[
\Delta a_t(H_t)
\{m_{t+1}(V,f_{t+1})-f_t(H_t)\}
\right]
=\E_t\left[
\Delta a_t(H_t)\E_t
\{m_{t+1}(V,f_{t+1})-f_t(H_t)|H_t\}
\right]
=0. \qed
$$
\end{enumerate}

\subsection{Proof of Corollary~\ref{cor:dml}}

We prove the result with the sample-splitting version, as in Algorithm~\ref{algo:dml}. For cross-fitting, simply apply the argument fold by fold, for a fixed number of folds.

Matching Algorithm~\ref{algo:dml}, we define the oracle analogues of the time-specific terms. Let
$
\phi_{0,i}=m_1(V_i,f_1)
$
and, for $t=1,\ldots,T$, let 
$
\phi_{t,i}
=
a_t(H_{t,i})
\{m_{t+1}(V_i,f_{t+1})-f_t(H_{t,i})\}. 
$
As a convention, take $m_{T+1}(V_i,f_{T+1})=Y_i$.

Next, we introduce notation for the proportion of the observations corresponding to each measure:
$
p_t=\E(R_{t,i}).
$
Since $n_t$ is of order $n$, we have $p_t$ bounded away from zero and $n_t/n\overset{p}{\rightarrow} p_t$.

We express the recursive parameter and the oracle influence in terms of proportions, to facilitate analysis at the rate $n^{-1/2}$. By Theorem~\ref{theorem:orthogonality},
$$
\theta_0=\sum_{t=0}^T \theta^{(t)},\quad
\theta^{(t)}=\E_t(\phi_t)=\frac{\E(R_{t,i}\phi_{t,i})}{p_t}.
$$
Moreover, the oracle influence of observation $i\in\test$ satisfies
$$
\psi_i
=
\sum_{t=0}^T
\frac{R_{t,i}}{p_t}
\{\phi_{t,i}-\theta^{(t)}\},\quad
\E(\psi_i)=0,\quad
\sigma^2=\E(\psi_i^2).
$$
By Assumption~\ref{assumption:variance}, the asymptotic variance $\sigma^2$  is finite. We proceed in steps.

\begin{enumerate}
    \item Consider the oracle version of the point estimator in Algorithm~\ref{algo:dml}:
    $$
    \bar{\theta}
    =
    \sum_{t=0}^T\bar{\theta}_t,\quad
    \bar{\theta}_t
    =
    \frac{1}{n_t}
    \sum_{i\in\test}R_{t,i}\phi_{t,i}.
    $$
    A standard expansion for a ratio estimator gives, for each $t$,
    $$
    n^{1/2}\{\bar{\theta}_t-\theta^{(t)}\}
    =
    n^{-1/2}
    \sum_{i\in\test}
    \frac{R_{t,i}}{p_t}
    \{\phi_{t,i}-\theta^{(t)}\}
    +
    o_p(1).
    $$
    Summing over $t=0,\ldots,T$, we conclude that
    $
    n^{1/2}(\bar{\theta}-\theta_0)
    =
    n^{-1/2}
    \sum_{i\in\test}\psi_i
    +
    o_p(1).
    $
    Since the $\test$ observations are independent and identically distributed, and $\sigma^2<\infty$, the central limit theorem gives
    $
    n^{1/2}(\bar{\theta}-\theta_0)
    \rightsquigarrow
    \mathcal{N}(0,\sigma^2).
    $

    \item Next, we show that the feasible estimator $\hat{\theta}$ is asymptotically equivalent to the oracle estimator $\bar{\theta}$. Define the empirical expectation
    $
    \hat{\E}_t(V)
    =
    \frac{1}{n_t}
    \sum_{i\in\test}R_{t,i}V_i.
    $
    Then
    $
    \hat{\theta}-\bar{\theta}
    =
    \sum_{t=0}^T
    \hat{\E}_t(\hat{\phi}_t-\phi_t).
    $
    
    To lighten notation, define
    $
    \Delta f_t=\hat f_t-f_t
    $
    and
    $
    \Delta a_t=\hat a_t-a_t.
    $
    By linearity of $f_t\mapsto m_t(V,f_t)$,
    $
    \hat{\phi}_0-\phi_0
    =
    m_1(V,\Delta f_1),
    $
    and, for $t=1,\ldots,T$,
    \begin{align*}
    \hat{\phi}_t-\phi_t
    &=
    a_t(H_t)
    \{m_{t+1}(V,\Delta f_{t+1})-\Delta f_t(H_t)\}\\
    &\quad+
    \Delta a_t(H_t)
    \{m_{t+1}(V,f_{t+1})-f_t(H_t)\}\\
    &\quad+
    \Delta a_t(H_t)
    \{m_{t+1}(V,\Delta f_{t+1})-\Delta f_t(H_t)\},
    \end{align*}
    with the convention $m_{T+1}(V,\Delta f_{T+1})=0$.
    Therefore
    \begin{align*}
    \hat{\theta}-\bar{\theta}
    &=
    \hat{\E}_0\{m_1(V,\Delta f_1)\}+
    \sum_{t=1}^T
    \hat{\E}_t\left[
    a_t(H_t)
    \{m_{t+1}(V,\Delta f_{t+1})-\Delta f_t(H_t)\}
    \right]\\
    &\quad+
    \sum_{t=1}^T
    \hat{\E}_t\left[
    \Delta a_t(H_t)
    \{m_{t+1}(V,f_{t+1})-f_t(H_t)\}
    \right]\\
    &\quad+
    \sum_{t=1}^T
    \hat{\E}_t\left[
    \Delta a_t(H_t)
    \{m_{t+1}(V,\Delta f_{t+1})-\Delta f_t(H_t)\}
    \right].
    \end{align*}

    \begin{enumerate}
\item We repeatedly use the following sample-splitting fact. For $i\in\test$ and $\hat{G}$ estimated from $\train$, define $\hat G(V_i)$. Conditional on $\train$, the function $\hat G$ is fixed over $\test$. Suppose that 
$
\E_t\{\hat G(V)| \train\}=0.
$

To begin, write the centered empirical process as 
$$
n^{1/2}\hat{\E}_t\{\hat G(V)\}
=
n^{1/2}\frac{1}{n_t}\sum_{i\in\test}R_{t,i}\hat G(V_i)
=
\frac{n}{n_t}\frac{1}{n^{1/2}}\sum_{i\in\test}R_{t,i}\hat G(V_i),
$$
Conditional on $\train$, the summands are independent and mean zero because
$$
\E\{R_{t,i}\hat G(V_i)|\train\}
=
p_t\E_t\{\hat G(V)|\train\}
=
0.
$$
Therefore, the empirical process has a second moment with vanishing cross terms:
\begin{align*}
&\E\left[
\left.
\left\{
\frac{1}{\sqrt n}
\sum_{i\in\test}R_{t,i}\hat G(V_i)
\right\}^2
\right|\train
\right]
 =
\frac{1}{n}
\sum_{i\in\test}
\E\left\{
R_{t,i}\hat G(V_i)^2
|\train
\right\} \\
&\quad =
p_t\E_t\{\hat G(V)^2|\train\}=p_t \|\hat G\|_t^2,
\end{align*}
where we slightly abuse notation by writing 
$
\|\hat G\|_t^2
=
\E_t\{\hat G(V)^2|\train\}.
$

Since $n_t/n\overset{p}{\rightarrow}p_t$ and $p_t$ is bounded away from zero, conditional Chebyshev's inequality bounds the centered empirical process as 
$
n^{1/2}\hat{\E}_t\{\hat G(V)\}
=
O_p\left(
\|\hat G\|_t
\right).
$
In particular, if $\|\hat G\|_t=o_p(1)$, then
$
n^{1/2}\hat{\E}_t\{\hat G(V)\}=o_p(1).
$

We also use the corresponding empirical-norm implication. If
$
\|\hat G\|_t=o_p(1),
$
then
$
\hat{\E}_t\{\hat G(V)^2\}=o_p(1).
$
Indeed, on the event $n_t/n\geq p_t/2$,
$$
\hat{\E}_t\{\hat G(V)^2\}
=
\frac{n}{n_t}\frac{1}{n}\sum_{i\in\test}R_{t,i}\hat G(V_i)^2
\leq
\frac{2}{p_t}\frac{1}{n}\sum_{i\in\test}R_{t,i}\hat G(V_i)^2.
$$
Taking  expectations conditional upon $\train$ gives
$$
\E\left\{
\left.
\frac{2}{p_t}\frac{1}{n}\sum_{i\in\test}R_{t,i}\hat G(V_i)^2
\right|\train
\right\}
=
2\E_t\{\hat G(V)^2|\train\}=2\|\hat{G}\|^2_t.
$$
The claim follows by conditional Markov's inequality.
    
        \item  By the same telescoping argument used in Lemma~\ref{lemma:bias},
    $$
    \E_0\{m_1(V,\Delta f_1)\}
    +
    \sum_{t=1}^T
    \E_t\left[
    a_t(H_t)
    \{m_{t+1}(V,\Delta f_{t+1})-\Delta f_t(H_t)\}
    \right]
    =
    0.
    $$
    Hence the first displayed line is a centered empirical-process. Conditional on $\train$, the functions $\hat f_t$ are fixed over $\test$. By Assumption~\ref{assumption:continuity}, 
    $
    \|m_1(V,\Delta f_1)\|_0
    \le
    \bar M\|\Delta f_1\|_1,
    $ 
    and, for $t=1,\ldots,T$, using the convention $\Delta f_{T+1}=0$,
    $$
    \left\|
    a_t(H_t)
    \{m_{t+1}(V,\Delta f_{t+1})-\Delta f_t(H_t)\}
    \right\|_t
    \le
    \bar a
    \left(
    \bar M\|\Delta f_{t+1}\|_{t+1}
    +
    \|\Delta f_t\|_t
    \right).
    $$
    Since $n_t$ is of order $n$, conditional Chebyshev's inequality and Assumption~\ref{assumption:rates} imply
    $$
    n^{1/2}
    \left|
    \hat{\E}_0\{m_1(V,\Delta f_1)\}
    +
    \sum_{t=1}^T
    \hat{\E}_t\left[
    a_t(H_t)
    \{m_{t+1}(V,\Delta f_{t+1})-\Delta f_t(H_t)\}
    \right]
    \right|
    =
    o_p(1).
    $$
        \item  Next, for each $t=1,\ldots,T$,
    $$
    \E_t\left[
    \Delta a_t(H_t)
    \{m_{t+1}(V,f_{t+1})-f_t(H_t)\}
    \right]
    =
    0
    $$
    by the law of iterated expectations and the recursive definition of $f_t$. Thus the second displayed line is also centered. Moreover,
    \begin{align*}
    &\left\|
    \Delta a_t(H_t)
    \{m_{t+1}(V,f_{t+1})-f_t(H_t)\}
    \right\|_t^2\\
    &\quad=
    \E_t\left[
    \Delta a_t(H_t)^2
    \E_t\left\{
    \{m_{t+1}(V,f_{t+1})-f_t(H_t)\}^2
    | H_t
    \right\}
    \right]\\
    &\quad\le
    \bar\sigma^2\|\Delta a_t\|_t^2,
    \end{align*}
    by Assumption~\ref{assumption:variance}. Since $n_t$ is of order $n$, conditional Chebyshev's inequality and Assumption~\ref{assumption:rates} imply
    $$
    n^{1/2}
    \left|
    \sum_{t=1}^T
    \hat{\E}_t\left[
    \Delta a_t(H_t)
    \{m_{t+1}(V,f_{t+1})-f_t(H_t)\}
    \right]
    \right|
    =
    o_p(1).
    $$
        \item The third displayed line is second order. By Cauchy--Schwarz,
    \begin{align*}
    &\left|
    \hat{\E}_t\left[
    \Delta a_t(H_t)
    \{m_{t+1}(V,\Delta f_{t+1})-\Delta f_t(H_t)\}
    \right]
    \right|\\
    &\quad\le
    \{\hat{\E}_t(\Delta a_t^2)\}^{1/2}
    \left\{
    \hat{\E}_t[
    \{m_{t+1}(V,\Delta f_{t+1})-\Delta f_t(H_t)\}^2]
    \right\}^{1/2}.
    \end{align*}
    Conditional on $\train$, the empirical norms have the same order as their population counterparts. Hence
    $
    \{\hat{\E}_t(\Delta a_t^2)\}^{1/2}
    =
    O_p(\|\Delta a_t\|_t),
    $
    and, by Assumption~\ref{assumption:continuity},
    $$
    \left\{
    \hat{\E}_t[
    \{m_{t+1}(V,\Delta f_{t+1})-\Delta f_t(H_t)\}^2]
    \right\}^{1/2}
    =
    O_p\left(
    \bar M\|\Delta f_{t+1}\|_{t+1}
    +
    \|\Delta f_t\|_t
    \right).
    $$
    Therefore, by Assumption~\ref{assumption:rates},
    $$
    n^{1/2}
    \left|
    \sum_{t=1}^T
    \hat{\E}_t\left[
    \Delta a_t(H_t)
    \{m_{t+1}(V,\Delta f_{t+1})-\Delta f_t(H_t)\}
    \right]
    \right|
    =
    o_p(1).
    $$
    \end{enumerate}

    Combining the three bounds, we obtain
    $
    n^{1/2}(\hat{\theta}-\bar{\theta})=o_p(1).
    $
    Therefore,
    $
    n^{1/2}(\hat{\theta}-\theta_0)
    \rightsquigarrow
    \mathcal{N}(0,\sigma^2).
    $
    Consistency follows immediately.

    \item What remains is to show consistency of the variance estimator. Consider the empirical influence notation
    $$
    \hat{\psi}_i
    =
    \sum_{t=0}^T
    \frac{n}{n_t}
    R_{t,i}
    \{\hat{\phi}_{t,i}-\hat{\theta}_t\},\quad   \bar{\psi}_i
    =
    \sum_{t=0}^T
    \frac{n}{n_t}
    R_{t,i}
    \{\phi_{t,i}-\bar{\theta}_t\}.
    $$
    
    \begin{enumerate}
        \item We first show that replacing $\phi_t$ by $\hat{\phi}_t$ is negligible:
$
\frac{1}{n}
\sum_{i\in\test}
(\hat{\psi}_i-\bar{\psi}_i)^2
=
o_p(1).
$
Since $T$ is fixed and $n/n_t=O_p(1)$, it is enough to control each time-specific component. For any fixed $t$,
\begin{align*}
\frac{1}{n}
\sum_{i\in\test}
\left[
\frac{n}{n_t}
R_{t,i}
\{(\hat{\phi}_{t,i}-\phi_{t,i})-(\hat{\theta}_t-\bar{\theta}_t)\}
\right]^2
&=
\frac{n}{n_t}
\hat{\E}_t\left[
\{(\hat{\phi}_t-\phi_t)-\hat{\E}_t(\hat{\phi}_t-\phi_t)\}^2
\right]\\
&\leq
\frac{n}{n_t}
\hat{\E}_t\{(\hat{\phi}_t-\phi_t)^2\}.
\end{align*}
Thus it suffices to show
$
\hat{\E}_t\{(\hat{\phi}_t-\phi_t)^2\}=o_p(1)
$
for each $t=0,\ldots,T$.

Throughout this step, we condition on $\train$, so the functions $\hat f_t$ and $\hat a_t$ are fixed over $\test$. We repeatedly use the following implication: if a function has population norm that is $o_p(1)$ under $\E_t$, then its subgroup empirical second moment under $\hat{\E}_t$ is also $o_p(1)$. This follows from independence of $\train$ and $\test$, the fact that $n_t$ is of order $n$, and conditional Markov's inequality.

For $t=0$,
$
\hat{\phi}_0-\phi_0=m_1(V,\Delta f_1).
$
By Assumption~\ref{assumption:continuity},
$
\|m_1(V,\Delta f_1)\|_0
\leq
\bar M\|\Delta f_1\|_1=o_p(1).
$
Therefore
$
\hat{\E}_0\{(\hat{\phi}_0-\phi_0)^2\}=o_p(1).
$

For $t=1,\ldots,T$, write
$$
\hat{\phi}_t-\phi_t
=
\Delta a_t(H_t)
\{m_{t+1}(V,f_{t+1})-f_t(H_t)\}
+
\hat a_t(H_t)
\{m_{t+1}(V,\Delta f_{t+1})-\Delta f_t(H_t)\}.
$$
In this variance-consistency step only, we use the boundedness of $\hat a_t$ from Assumption~\ref{assumption:variance}.

For the first term, Assumption~\ref{assumption:variance} gives
\begin{align*}
\left\|
\Delta a_t(H_t)
\{m_{t+1}(V,f_{t+1})-f_t(H_t)\}
\right\|_t^2
&=
\E_t\left[
\Delta a_t(H_t)^2
\E_t\left\{
\{m_{t+1}(V,f_{t+1})-f_t(H_t)\}^2
| H_t
\right\}
\right]\\
&\leq
\bar\sigma^2\|\Delta a_t\|_t^2
=
o_p(1).
\end{align*}
Hence its empirical second moment is $o_p(1)$.

For the second term, boundedness of $\hat a_t$ and Assumption~\ref{assumption:continuity} imply
$$\left\|
\hat a_t(H_t)
\{m_{t+1}(V,\Delta f_{t+1})-\Delta f_t(H_t)\}
\right\|_t
\leq
\bar a
\left(
\bar M\|\Delta f_{t+1}\|_{t+1}
+
\|\Delta f_t\|_t
\right)
=
o_p(1)$$
with the convention $m_{T+1}(V,\Delta f_{T+1})=0$. Hence its empirical second moment is also $o_p(1)$.

Combining the two terms gives
$
\hat{\E}_t\{(\hat{\phi}_t-\phi_t)^2\}=o_p(1)
$
for each $t=0,\ldots,T$. Therefore
$
\frac{1}{n}
\sum_{i\in\test}
(\hat{\psi}_i-\bar{\psi}_i)^2
=
o_p(1).
$
       \item  Next, we show
        $
        \frac{1}{n}
        \sum_{i\in\test}\bar{\psi}_i^2
        =
        \frac{1}{n}
        \sum_{i\in\test}\psi_i^2
        +
        o_p(1).
        $
        
        To begin, we show
        $
        \frac{1}{n}
        \sum_{i\in\test}
        (\bar{\psi}_i-\psi_i)^2
        =
        o_p(1).
        $
        Since $T$ is fixed, it suffices to verify this for each time-specific component. For any fixed $t$,
        \begin{align*}
        &\frac{1}{n}
        \sum_{i\in\test}
        \left[
        \frac{n}{n_t}
        R_{t,i}\{\phi_{t,i}-\bar{\theta}_t\}
        -
        \frac{1}{p_t}
        R_{t,i}\{\phi_{t,i}-\theta^{(t)}\}
        \right]^2\\
        &\quad =
        \frac{1}{n}
        \sum_{i\in\test}
        \left[
        \left(\frac{n}{n_t}-\frac{1}{p_t}\right)
        R_{t,i}\{\phi_{t,i}-\theta^{(t)}\}
        -
        \frac{n}{n_t}
        R_{t,i}\{\bar{\theta}_t-\theta^{(t)}\}
        \right]^2\\
        &\quad \leq
        2\left(\frac{n}{n_t}-\frac{1}{p_t}\right)^2
        \frac{1}{n}
        \sum_{i\in\test}
        R_{t,i}\{\phi_{t,i}-\theta^{(t)}\}^2+
        2\left(\frac{n}{n_t}\right)^2
        \{\bar{\theta}_t-\theta^{(t)}\}^2
        \frac{1}{n}
        \sum_{i\in\test}R_{t,i}.
        \end{align*}
        The first term is $o_p(1)$ because $n_t/n\overset{p}{\rightarrow}p_t$ and
        $
        \frac{1}{n}
        \sum_{i\in\test}
        R_{t,i}\{\phi_{t,i}-\theta^{(t)}\}^2
        =
        O_p(1).
        $
        The second term is $o_p(1)$ because $n/n_t=O_p(1)$,
        $\bar{\theta}_t\overset{p}{\rightarrow}\theta^{(t)}$, and
        $n^{-1}\sum_{i\in\test}R_{t,i}=O_p(1)$. Hence
        $
        \frac{1}{n}
        \sum_{i\in\test}
        (\bar{\psi}_i-\psi_i)^2
        =
        o_p(1)
        $ as desired.

     Using $u^2-v^2=(v+u-v)^2-v^2=(u-v)^2+2(u-v)v$ and Cauchy-Schwarz, 
        \begin{align*}
        \left|
        \frac{1}{n}
        \sum_{i\in\test}\bar{\psi}_i^2
        -
        \frac{1}{n}
        \sum_{i\in\test}\psi_i^2
        \right|
        &\leq
        \frac{1}{n}
        \sum_{i\in\test}(\bar{\psi}_i-\psi_i)^2
        +
        2
        \left\{
        \frac{1}{n}
        \sum_{i\in\test}(\bar{\psi}_i-\psi_i)^2
        \right\}^{1/2}
        \left\{
        \frac{1}{n}
        \sum_{i\in\test}\psi_i^2
        \right\}^{1/2}.
        \end{align*}
        By the law of large numbers,
        $
        \frac{1}{n}
        \sum_{i\in\test}\psi_i^2
        \overset{p}{\rightarrow}
        \E(\psi_i^2)
        =
        \sigma^2.
        $
        Thus
        $
        \frac{1}{n}
        \sum_{i\in\test}\bar{\psi}_i^2
        =
        \sigma^2+o_p(1).
        $
    \end{enumerate}

    We combine the two variance approximations to prove
    $
    \hat{\sigma}^2
    =
    \frac{1}{n}
    \sum_{i\in\test}\hat{\psi}_i^2
    =
    \sigma^2+o_p(1).
    $
  Using $u^2-v^2=(v+u-v)^2-v^2=(u-v)^2+2(u-v)v$ and Cauchy-Schwarz, 
    \begin{align*}
    \left|
    \frac{1}{n}\sum_{i\in\test}\hat{\psi}_i^2
    -
    \frac{1}{n}\sum_{i\in\test}\bar{\psi}_i^2
    \right|
    &\leq
    \frac{1}{n}\sum_{i\in\test}(\hat{\psi}_i-\bar{\psi}_i)^2
    +
    2
    \left\{
    \frac{1}{n}\sum_{i\in\test}(\hat{\psi}_i-\bar{\psi}_i)^2
    \right\}^{1/2}
    \left\{
    \frac{1}{n}\sum_{i\in\test}\bar{\psi}_i^2
    \right\}^{1/2}
    =o_p(1),
    \end{align*}
    since we have shown 
    $
    \frac{1}{n}\sum_{i\in\test}(\hat{\psi}_i-\bar{\psi}_i)^2=o_p(1)
    $
     and
    $
    \frac{1}{n}\sum_{i\in\test}\bar{\psi}_i^2=O_p(1).
    $
    
    Finally, Slutsky's theorem gives
    $
    \P\left(
    \theta_0\in
    [\hat{\theta}\pm 1.96 n^{-1/2}\hat{\sigma}]
    \right)
    \rightarrow
    0.95.
    $ \qed
\end{enumerate}

\section{Proofs for Section~\ref{sec:riesz}}\label{sec:riesz_proof}

\subsection{Critical radius theory}

We recap critical radius theory used to prove Theorem~\ref{theorem:rate}, quantifying complexity in a measure-specific way. For each $t$, the empirical expectation
$
\hat{\E}_t\{g(V)\}
=
\frac{1}{n_t}\sum_{i\in\train}R_{t,i}g(V_i)
$
averages over the subgroup corresponding to $\E_t(\cdot)$. Thus the norm, Rademacher complexity, and critical radius depend on the measure $\E_t(\cdot)$. 

We have already introduced notation for the measure-specific norm: 
$
\|g\|_t^2=\E_t\{g(V)^2\}.
$

Now, we introduce notation for the measure-specific Rademacher complexity. For a class $\mathcal G$ evaluated under $\E_t$, its local Rademacher complexity is
$$
\mathcal R_t(\delta;\mathcal G)
=
\E\left[
\sup_{g\in\mathcal G:\ \E_t\{g(V)^2\}\leq \delta^2}
\left|
\frac{1}{n_t}\sum_{i\in\train}R_{t,i}\varepsilon_i g(V_i)
\right|
\right],
$$
where $\varepsilon_i$ are independent Rademacher random variables. 
Intuitively, $\mathcal R_t(\delta;\mathcal G)$ is a localized measure of the complexity of $\mathcal G$: it measures how well functions norm at most $\delta$ can fit pure noise. Through Dudley's entropy integral, this quantity is controlled by covering numbers under the measure $\E_t(\cdot)$. For VC-type classes, it is governed by the VC dimension up to logarithmic factors.

Finally, we introduce notation for the measure-specific critical radius.
A critical radius $\delta_t$ is any sequence satisfying
$
\mathcal R_t(\delta_t;\mathcal G)\leq \delta_t^2.
$
In our setting, it is measure-specific: the same algebraic class can have different critical radii under $\E_t(\cdot)$ and $\E_{t-1}(\cdot)$ because both the norm and the sampling distribution change. 
Intuitively, the critical radius is the estimation scale at which the random fluctuation of the empirical criterion is dominated by the quadratic curvature of the population criterion. Since the fluctuation is of order $\mathcal R_t(\delta;\mathcal G)$ and the curvature is of order $\delta^2$, the inequality $\mathcal R_t(\delta_t;\mathcal G)\leq \delta_t^2$ yields the mean-square rate. For example, in a $d$-dimensional or VC-type class, this gives $\delta_t^2$ of order $d/n_t$, up to logarithmic factors.

We use a standard concentration result in terms of the critical radius.

\begin{lemma}[Critical radius concentration]\label{lemma:critical-radius-concentration}
Fix a measure $\E_t$, sample size $n_t$, and a class $\mathcal G$ of uniformly bounded functions. Let $g_0\in\mathcal G$, and suppose
$
\operatorname{star}(\mathcal G-g_0)
=
\{\kappa(g-g_0):g\in\mathcal G,\ \kappa\in[0,1]\}
$
has critical radius $\delta_t$ under $\E_t(\cdot)$.
Let $\ell(u,v)$ be Lipschitz in $u$ with Lipschitz constant $L$. If
$
\delta_t^2
\geq
c\frac{\ln\{\ln(n_t)\}}{n_t}
$
for a universal constant $c>0$, then with probability at least $1-\eta$, uniformly over $g\in\mathcal G$,
$$
\begin{aligned}
&\left|
(\hat{\E}_t-\E_t)
\left[
\ell\{g(V),V\}
-
\ell\{g_0(V),V\}
\right]
\right|
\leq
CL
\left(
\delta_t+\sqrt{\frac{\ln(1/\eta)}{n_t}}
\right)
\|g-g_0\|_t
+
CL^2
\left(
\delta_t^2+\frac{\ln(1/\eta)}{n_t}
\right),
\end{aligned}
$$
where $C<\infty$ is universal.
If $\ell(u,v)$ is linear in $u$, the lower bound on $\delta_t^2$ is not needed.
\end{lemma}

\begin{proof}
    The result is an immediate extension of \citet[Lemma 14]{foster2023orthogonal}.
\end{proof}

We apply Lemma~\ref{lemma:critical-radius-concentration} to the two empirical-process terms that appear in the recursive Riesz loss. Recall the current-measure class
$$
\mathcal F_t
=
\left[
h_t\mapsto \kappa\{\tilde a_t(h_t)-a_t(h_t)\}:
\tilde a_t\in\mathcal A,\ \kappa\in[0,1]
\right],
$$
evaluated under $\E_t(\cdot)$, with critical radius be $\delta_t$. Recall the previous-measure class
$$
\mathcal F_t'
=
\left[
v\mapsto
\kappa\,\tilde a_{t-1}(h_{t-1})m_t(v,\tilde a_t-a_t):
\tilde a_t,\tilde a_{t-1}\in\mathcal A,\ \kappa\in[0,1]
\right],
$$
evaluated under $\E_{t-1}(\cdot)$, with critical radius be $\delta_t'$. As a convention, take $\tilde a_0=a_0=1$.

\begin{corollary}[Critical-radius bounds for the recursive Riesz loss]\label{cor:critical-radius-riesz}
Suppose the conditions of Theorem~\ref{theorem:rate} hold. Then with probability at least $1-\eta$, for every $t=1,\ldots,T$ and every $\tilde a_t,\tilde a_{t-1}\in\mathcal A$,
\begin{align*}
\left|
(\hat{\E}_t-\E_t)
\{\tilde a_t(H_t)^2-a_t(H_t)^2\}
\right|
&\leq
\frac{1}{8}\|\tilde a_t-a_t\|_t^2
+
C\bar a^2
\left\{
\delta_t^2+\frac{\ln(T/\eta)}{n_t}
\right\}\\
\left|
(\hat{\E}_{t-1}-\E_{t-1})
\{\tilde a_{t-1}(H_{t-1})m_t(V,\tilde a_t-a_t)\}
\right|
&\leq
\frac{1}{8}\|\tilde a_t-a_t\|_t^2
+
C(\bar a\bar M)^2
\left\{
(\delta_t')^2+\frac{\ln(T/\eta)}{n_{t-1}}
\right\}.
\end{align*}
Here $C<\infty$ is universal.
\end{corollary}

\begin{proof}[Proof of Corollary~\ref{cor:critical-radius-riesz}]
For the first display, write
$$
\tilde a_t(H_t)^2-a_t(H_t)^2
=
\{\tilde a_t(H_t)-a_t(H_t)\}
\{\tilde a_t(H_t)+a_t(H_t)\}.
$$
Since $\mathcal A$ is uniformly bounded by $\bar a$, the map $u\mapsto u^2$ is $2\bar a$-Lipschitz on the range of $\mathcal A$. Applying Lemma~\ref{lemma:critical-radius-concentration} to $\mathcal F_t$ with failure probability $\eta/(2T)$ gives
$$
\left|
(\hat{\E}_t-\E_t)
\{\tilde a_t(H_t)^2-a_t(H_t)^2\}
\right|
\leq
C\bar a
\left(
\delta_t+\sqrt{\frac{\ln(2T/\eta)}{n_t}}
\right)
\|\tilde a_t-a_t\|_t
+
C\bar a^2
\left(
\delta_t^2+\frac{\ln(2T/\eta)}{n_t}
\right).
$$
Using $2uv\leq u^2/8+C v^2$ gives the desired result.

For the second display, Assumption~\ref{assumption:continuity} and boundedness of $\tilde a_{t-1}$ imply
$$
\|\tilde a_{t-1}(H_{t-1})m_t(V,\tilde a_t-a_t)\|_{t-1}
\leq
\bar a\bar M\|\tilde a_t-a_t\|_t.
$$
The empirical-process term is linear in the class $\mathcal F_t'$. Applying the linear part of Lemma~\ref{lemma:critical-radius-concentration} with failure probability $\eta/(2T)$ gives
\begin{align*}
    &\left|
(\hat{\E}_{t-1}-\E_{t-1})
\{\tilde a_{t-1}(H_{t-1})m_t(V,\tilde a_t-a_t)\}
\right|\\
&
\leq
C
\left\{
\delta_t'+\sqrt{\frac{\ln(2T/\eta)}{n_{t-1}}}
\right\}
\bar a\bar M\|\tilde a_t-a_t\|_t
+
C(\bar a\bar M)^2
\left\{
(\delta_t')^2+\frac{\ln(2T/\eta)}{n_{t-1}}
\right\}.
\end{align*}
Using $2uv\leq u^2/8+C v^2$ and $\bar a,\bar M\geq1$ yields the desired result.

The probability statement follows by taking a union bound over $t=1,\ldots,T$.
\end{proof}

\subsection{Proof of Theorem~\ref{theorem:rate}}

Fix $t\in\{1,\ldots,T\}$. To lighten notation, define
$
\Delta a_t=\hat a_t-a_t
$
and
$
\Delta a_{t-1}=\hat a_{t-1}-a_{t-1}.
$ Since we have fixed $t$, we can abbreviate the placeholder as $a=\tilde{a}_t$.

Define the population Riesz loss
$$
\mathcal L_t(a)
=
\E_t\{a(H_t)^2\}
-
2\E_{t-1}\{a_{t-1}(H_{t-1})m_t(V,a)\},
$$
and the empirical Riesz loss used by Algorithm~\ref{algo:riesz},
$$
\hat{\mathcal L}_t(a)
=
\hat{\E}_t\{a(H_t)^2\}
-
2\hat{\E}_{t-1}\{\hat a_{t-1}(H_{t-1})m_t(V,a)\}.
$$

\begin{enumerate}
    \item We record the population curvature. 
   Using Assumption~\ref{assumption:existence} and the identity $u^2-v^2=(v+u-v)^2-v^2=(u-v)^2+2(u-v)v$, 
    \begin{align*}
    \mathcal L_t(a)-\mathcal L_t(a_t)
    &=
    \E_t\{a(H_t)^2-a_t(H_t)^2\}
    -
    2\E_{t-1}\{a_{t-1}(H_{t-1})m_t(V,a-a_t)\}\\
    &=
    \E_t\{a(H_t)^2-a_t(H_t)^2\}
    -
    2\E_t\{a_t(H_t)(a-a_t)(H_t)\}\\
    &=
    \|a-a_t\|_t^2.
    \end{align*}
    In particular,
    $
    \mathcal L_t(\hat a_t)-\mathcal L_t(a_t)
    =
    \|\Delta a_t\|_t^2.
    $

    \item We now use empirical minimization. 

By the definition of the empirical Riesz loss and the linearity of
$a\mapsto m_t(V,a)$,
$$
\hat{\mathcal L}_t(\hat a_t)-\hat{\mathcal L}_t(a_t)
=
\hat{\E}_t\{\hat a_t(H_t)^2-a_t(H_t)^2\}
-
2\hat{\E}_{t-1}\{\hat a_{t-1}(H_{t-1})m_t(V,\Delta a_t)\}.
$$
Add and subtract the corresponding population terms:
\begin{align*}
    \hat{\E}_t\{\hat a_t(H_t)^2-a_t(H_t)^2\}
&=
\E_t\{\hat a_t(H_t)^2-a_t(H_t)^2\}
+
(\hat{\E}_t-\E_t)\{\hat a_t(H_t)^2-a_t(H_t)^2\} \\
\hat{\E}_{t-1}\{\hat a_{t-1}(H_{t-1})m_t(V,\Delta a_t)\}
&=
\E_{t-1}\{a_{t-1}(H_{t-1})m_t(V,\Delta a_t)\}\\
&\quad+
(\hat{\E}_{t-1}-\E_{t-1})
\{\hat a_{t-1}(H_{t-1})m_t(V,\Delta a_t)\}\\
&\quad+
\E_{t-1}\{\Delta a_{t-1}(H_{t-1})m_t(V,\Delta a_t)\}.
\end{align*}
Substituting these decompositions gives
\begin{align*}
\hat{\mathcal L}_t(\hat a_t)-\hat{\mathcal L}_t(a_t)
&=
\E_t\{\hat a_t(H_t)^2-a_t(H_t)^2\}
-
2\E_{t-1}\{a_{t-1}(H_{t-1})m_t(V,\Delta a_t)\}\\
&\quad+
(\hat{\E}_t-\E_t)\{\hat a_t(H_t)^2-a_t(H_t)^2\}\\
&\quad-
2(\hat{\E}_{t-1}-\E_{t-1})
\{\hat a_{t-1}(H_{t-1})m_t(V,\Delta a_t)\}\\
&\quad-
2\E_{t-1}\{\Delta a_{t-1}(H_{t-1})m_t(V,\Delta a_t)\}.
\end{align*}
By the previous step, 
$$
\E_t\{\hat a_t(H_t)^2-a_t(H_t)^2\}
-
2\E_{t-1}\{a_{t-1}(H_{t-1})m_t(V,\Delta a_t)\}=\|\Delta a_t\|_t^2.
$$

Since $\hat a_t$ minimizes $\hat{\mathcal L}_t$ over $\mathcal A$ and $a_t\in\mathcal A$,
    $
    \hat{\mathcal L}_t(\hat a_t)-\hat{\mathcal L}_t(a_t)\leq 0,
    $ and hence 
\begin{align*}
\|\Delta a_t\|_t^2
&\leq
\underbrace{\left|
(\hat{\E}_t-\E_t)\{\hat a_t(H_t)^2-a_t(H_t)^2\}
\right|
+
2\left|
(\hat{\E}_{t-1}-\E_{t-1})
\{\hat a_{t-1}(H_{t-1})m_t(V,\Delta a_t)\}
\right|}_{\text{empirical process}}\\
&\quad+
\underbrace{2\left|
\E_{t-1}\{\Delta a_{t-1}(H_{t-1})m_t(V,\Delta a_t)\}
\right|}_{\text{recursive error}}.
\end{align*}

    \item We control the empirical-process terms. By Corollary~\ref{cor:critical-radius-riesz}, with probability at least $1-\eta$,
\begin{align*}
&\left|
(\hat{\E}_t-\E_t)\{\hat a_t(H_t)^2-a_t(H_t)^2\}
\right|
+
2\left|
(\hat{\E}_{t-1}-\E_{t-1})
\{\hat a_{t-1}(H_{t-1})m_t(V,\Delta a_t)\}
\right|\\
&\leq
\frac{1}{2}\|\Delta a_t\|_t^2
+
C(\bar a\bar M)^2
\left[
\max(\delta_t,\delta_t')^2
+
\frac{\ln(T/\eta)}{\min(n_t,n_{t-1})}
\right].
\end{align*}

    \item We control the recursive error term. By Cauchy--Schwarz and Assumption~\ref{assumption:continuity},
    \begin{align*}
    &2\left|
    \E_{t-1}\{\Delta a_{t-1}(H_{t-1})m_t(V,\Delta a_t)\}
    \right|
    \leq
    2\|\Delta a_{t-1}\|_{t-1}
    \|m_t(V,\Delta a_t)\|_{t-1}\\
    &\quad\leq
    2\bar M\|\Delta a_{t-1}\|_{t-1}\|\Delta a_t\|_t
    \leq
    \frac{1}{4}\|\Delta a_t\|_t^2
    +
    C\bar M^2\|\Delta a_{t-1}\|_{t-1}^2.
    \end{align*}
    Since $\bar a,\bar M\geq 1$, the final term is bounded by
    $
    C(\bar a\bar M)^2\|\Delta a_{t-1}\|_{t-1}^2.
    $

    \item Combining the empirical-process bounds and the recursive error bound, we obtain
    $$
    \|\Delta a_t\|_t^2
    \leq
    \frac{3}{4}\|\Delta a_t\|_t^2
    +
    C(\bar a\bar M)^2
    \left\{
    \max(\delta_t,\delta_t')^2
    +
    \frac{\ln(T/\eta)}{\min(n_t,n_{t-1})}
    +
    \|\Delta a_{t-1}\|_{t-1}^2
    \right\}.
    $$
    Moving the first term on the right-hand side to the left-hand side and enlarging the universal constant $C$ gives
    $$
    \|\hat a_t-a_t\|_t^2
    \leq
    C(\bar a\bar M)^2
    \left\{
    \max(\delta_t,\delta_t')^2
    +
    \frac{\ln(T/\eta)}{\min(n_t,n_{t-1})}
    +
    \|\hat a_{t-1}-a_{t-1}\|_{t-1}^2
    \right\}.
    $$
    A union bound over $t=1,\ldots,T$ gives the one-step result  simultaneously for every time period, with total probability at least $1-\eta$.
\item Finally, we iterate the recursion. Write the one-step result as
$$
  B_t
    \leq
    K(R_t+B_{t-1}),
    \quad  B_t=\|\hat a_t-a_t\|_t^2,\quad  R_t=
    \max(\delta_t,\delta_t')^2
    +
    \frac{\ln(T/\eta)}{\min(n_t,n_{t-1})},\quad 
    K=C(\bar a\bar M)^2
$$
    with $B_0=0$. Enlarge the universal constant $C$ if necessary so that $K\geq 1$. Iterating the recursion,
    $
    B_t
    \leq
    \sum_{s=1}^t K^{t-s+1}R_s.
    $
    Since $t\leq T$,
    $
    B_t
    \leq
    T K^T\max_{1\leq s\leq T}R_s.
    $
    Substituting the definition of $R_s$ gives
    $$
    \|\hat a_t-a_t\|_t^2
    \leq
    T\{C(\bar a\bar M)^2\}^T
    \left[
    \max_{s\in\{1,\ldots,T\}}
    \left\{
    \max(\delta_s,\delta_s')^2
    \right\}
    +
    \frac{\ln(T/\eta)}{\min_{s\in\{0,\ldots,T\}}n_s}
    \right]. \qed
    $$

\end{enumerate}

\clearpage

\begin{center}
{\LARGE\bf Online Appendix}
\end{center}

Appendix~\ref{sec:nonlinear} extends our framework to nonlinear recursive functionals, with a leading example of dynamic discrete choice.
Appendix~\ref{sec:instrument} extends our framework to instrumental variables.
Appendix~\ref{sec:clever} formalizes the connection to clever covariates.
Appendices~\ref{sec:sim} and~\ref{sec:app} provide details for our simulations and applications, respectively.

\section{Extension to nonlinear recursive functionals}\label{sec:nonlinear}

The main text focuses on a rich class of recursive functionals with three properties: (i) a scalar target parameter $\theta_0$, (ii) explicit identification of the form $\theta_0=\E_0\{m_1(V,f_1)\}$, and (iii) linearity of the time-varying formulas $f_t\mapsto m_t(V,f_t)$. While this class already contains the various examples of Section~\ref{sec:examples}, we now enrich it to cover some additional models in structural econometrics.

We generalize all three properties: (i) a vector target parameter $\theta_0$, (ii) implicit identification of the form $
\E_0\{m_1(V,f_1,\theta_0)\}=0
$, and (iii) nonlinearity of the time-varying formulas. 
The key insight is that a nonlinear yet differentiable time-varying formula is well approximated by its linear derivative.

We generalize our main result in Appendix~\ref{sec:derivative}. We recap a canonical dynamic discrete choice model in Appendix~\ref{sec:choice}. We apply our result to the  canonical model in Appendix~\ref{sec:match}. Finally, we describe further generalizations in Appendix~\ref{sec:extensions}. We defer proofs to the end.

\subsection{A generalized class}\label{sec:derivative}

We extend our main result to the following class of nonlinear recursive functionals.

\begin{definition}[Nonlinear recursive functional]\label{def:nonlinear}
Let $V$ concatenate all observed variables across time periods. Let $T\geq 1$ be fixed. For each time period $t$, let $H_t$ be the time-varying conditioning variables and let $\E_t(\cdot)$ be the expectation with respect to the time-varying measure.
We study the parameter $\theta_0$ implicitly defined by
$
\E_0\{m_1(V,f_1,\theta_0)\}=0.
$
As before, the nuisance functions are recursively defined by
$$
f_t(H_t)=\E_t\{m_{t+1}(V,f_{t+1})\mid H_t\},\quad t=1,\ldots,T-1,\quad f_T(H_T)=\E_T(Y\mid H_T).$$
\end{definition}

First, we allow the target parameter $\theta_0$ to be a vector. For simplicity, we impose that it is exactly identified, and the regressions $f_t$ are scalar-valued. The resulting Riesz representers $a_t$ are vector valued, with the same dimension as $\theta_0$. The notation remains essentially the same, and our analysis goes through component-wise. Further generalizations are possible, as discussed in Appendix~\ref{sec:extensions}, but unnecessary for the dynamic discrete choice model below.

Second, we implicitly define the parameter by $\E_0\{m_1(V,f_1,\theta_0)\}=0.$ For simplicity, we impose that $\theta_0$ appears in $m_1$ but not $m_t$ for $t\geq 2$. This restriction is satisfied by the dynamic discrete choice model below, and it lightens notation since it means that $f_t$ is not indexed by the parameter. The further extension is possible but unnecessary for our example.

Third, and most importantly, we allow the time-varying formulas $f_t\mapsto m_t(V,f_t)$ to be nonlinear, relaxing a key restriction in Definition~\ref{def:linear}. While the formulas may be nonlinear, we impose that they remain differentiable, in a manner that we now formalize.

\begin{assumption}[Differentiability and existence]\label{assumption:nonlinear-existence}
Consider the notation of Definition~\ref{def:nonlinear}. As a convention, let $a_0=1$. 
For each $t=1,\ldots,T$, the time-varying formula is differentiable; formally, for each placeholder function $\tilde{f}_t(H_t)$, the following pathwise derivative exists:\footnote{For $t=1$, the formula $m_1$ is evaluated at the true parameter value $\theta_0$.}
$$
\dot{m}_t(\tilde{f}_t)
=
\left.
\frac{d}{d\kappa}
\E_{t-1}\{a_{t-1}(H_{t-1})m_t(V,f_t+\kappa\tilde{f}_t)\}
\right|_{\kappa=0}.
$$
Moreover, there exists a Riesz representer $a_t(H_t)$ with finite variance satisfying
$$
\E_t\{a_t(H_t)\tilde{f}_t(H_t)\}
=
\dot{m}_t(\tilde{f}_t)
$$
for all placeholder functions $\tilde{f}_t(H_t)$ with finite variance.
\end{assumption}

Assumption~\ref{assumption:nonlinear-existence} is the nonlinear analogue of Assumption~\ref{assumption:existence}. In the linear case,
$
\dot{m}_t(\tilde{f}_t)
=
\E_{t-1}\{a_{t-1}(H_{t-1})m_t(V,\tilde{f}_t)\},
$
so Assumption~\ref{assumption:nonlinear-existence} reduces to Assumption~\ref{assumption:existence}. 

More generally, Assumption~\ref{assumption:nonlinear-existence} has two conditions,  which we verify for dynamic discrete choice models in Appendix~\ref{sec:match}. First, it requires that the nonlinear functional  $f_t\mapsto\E_{t-1}\{a_{t-1}(H_{t-1})m_t(V,f_t\}$  is smooth, which is satisfied by random utility models with type-I extreme value shocks. Second, it requires that the linear functional $\dot{m}_t(\tilde{f}_t)$, which locally approximates the nonlinear functional $f_t\mapsto\E_{t-1}\{a_{t-1}(H_{t-1})m_t(V,f_t\}$, has a Riesz representer $a_t$. This requirement is satisfied under standard overlap conditions discussed in Appendix~\ref{sec:match}.

Under these regularity conditions, which are standard in structural econometrics, our main result extends to nonlinear recursive functionals.

\begin{theorem}[Nonlinear recursive orthogonality]\label{theorem:nonlinear-orthogonality}
Consider the notation of Definition~\ref{def:nonlinear}. Suppose Assumption~\ref{assumption:nonlinear-existence} holds. Define
$$
\mathcal{M}(\tilde{f}_1,\ldots,\tilde{f}_T,\tilde{a}_1,\ldots,\tilde{a}_T,\theta)\\
=
\E_0\{m_1(V,\tilde{f}_1,\theta)\}
+
\sum_{t=1}^T
\E_t\left[
\tilde{a}_t(H_t)
\{m_{t+1}(V,\tilde{f}_{t+1})-\tilde{f}_t(H_t)\}
\right],
$$
with the convention $m_{T+1}(V,\tilde{f}_{T+1})=Y$. Then
$
\mathcal{M}(f_1,\ldots,f_T,a_1,\ldots,a_T,\theta_0)=0
$
is a Neyman orthogonal estimating equation. Here, $f_t$ is recursively defined from $f_{t+1}$ in Definition~\ref{def:nonlinear}, and $a_t$ is recursively defined from $a_{t-1}$ in Assumption~\ref{assumption:nonlinear-existence}.
\end{theorem}

We give the proof below, exactly generalizing the proof of Theorem~\ref{theorem:orthogonality}. 
The intuition remains the same: the regressions recurse forward in time, in that $f_t$ is defined from $f_{t+1}$, while the Riesz representers recurse backwards in time, in that $a_t$ is defined from $a_{t-1}$. 
As before, our contribution is a general construction, that is automatic in nature, in terms of recursive Riesz representers. 

In the same way that the orthogonality in Theorem~\ref{theorem:orthogonality} delivers debiased inference in Corollary~\ref{cor:dml}, the orthogonality in Theorem~\ref{theorem:nonlinear-orthogonality} implies debiased inference. 
As before, the orthogonal construction is new; thereafter, the asymptotic theory follows from earlier work.
For brevity, we do not write out the debiased inference corollary. See e.g. \cite{chernozhukov2022locally} for regularity conditions.\footnote{The bias expansion contains higher-order remainders from the nonlinear time-varying formulas, so inference requires additional smoothness and rate conditions.}

\subsection{Dynamic discrete choice}\label{sec:choice}

Consider a two-period binary dynamic discrete choice model. In each period, the agent observes a state $X_t$ and chooses an action $A_t\in\{0,1\}$. Flow payoffs are
$
R_t^{(j)}=d^{(j)}(X_t)^\top \theta_0+\varepsilon_t^{(j)}
$
for $j\in\{0,1\}$. To lighten notation, define the payoff-index difference
$
\Delta(X_t)=d^{(1)}(X_t)-d^{(0)}(X_t).
$
Assume the random utility shocks $\{\varepsilon_t^{(0)},\varepsilon_t^{(1)}\}$ are independent and identically distributed type-I extreme value. Then the shock difference has the logit distribution
$
\Lambda(s)=\frac{\exp(s)}{1+\exp(s)}.
$

The action $A_t=1$ means renewal, which has two implications. First, $d^{(1)}(X_t)$ is constant, so the renewal payoff is state-independent. Second, renewal resets the relevant state, so the continuation value after renewal does not depend on the current state.

We solve the model by backward induction. Consider the terminal time period $t=2$, and let $v_2^{(1)}(X_2)$ and $v_2^{(0)}(X_2)$ denote the terminal choice-specific values before shocks. The terminal choice maximizes utility after shocks:
$
A_2=1\{v_2^{(1)}(X_2)+\varepsilon_2^{(1)}\geq v_2^{(0)}(X_2)+\varepsilon_2^{(0)}\}.
$
Define the reduced form conditional choice probability
$
f_2(X_2)=\E(A_2\mid X_2).
$
Then
$
f_2(X_2)=\Lambda\{v_2^{(1)}(X_2)-v_2^{(0)}(X_2)\}.
$
By invertibility of the logit link,
$
v_2^{(1)}(X_2)-v_2^{(0)}(X_2)=\Lambda^{-1}\{f_2(X_2)\}.
$

At time period $t=1$, the agent forms an expectation of the maximized utility at $t=2$. The Hotz--Miller inversion is a way to characterize this expected maximum. Formally, define the expected terminal value by
$
V_2(X_2)
=
\E_{\varepsilon}\left[
\max_{j\in\{0,1\}}\{v_2^{(j)}(X_2)+\varepsilon_2^{(j)}\}
\mid X_2
\right].
$
Normalize relative to renewal. Then
$
V_2(X_2)
=
v_2^{(1)}(X_2)+Q\{f_2(X_2)\},
$
where
$
Q(s)=\gamma_E-\ln(s),
$
and $\gamma_E$ is Euler's constant. In summary, we have expressed the expected maximum as a known nonlinear function of the reduced form conditional choice probability.

Because the renewal payoff is state-independent, the $v_2^{(1)}(X_2)$ component of the terminal value cancels from the first-period continuation-value difference. Therefore the relevant continuation-value component is
$
f_1(X_1,A_1)
=
\E\{Q(f_2(X_2))\mid X_1,A_1\}.
$
The renewal restriction implies that $f_1(X_1,1)$ is constant in $X_1$, but we keep the notation $f_1(X_1,A_1)$ for symmetry.

Therefore, at $t=1$, the agent chooses
$$
A_1
=
1\{u(V,f_1,\theta_0)+\varepsilon_1^{(1)}-\varepsilon_1^{(0)}\geq 0\},
\quad 
u(V,\tilde{f}_1,\theta)
=
\Delta(X_1)^\top\theta
+
\beta\{\tilde{f}_1(X_1,1)-\tilde{f}_1(X_1,0)\},
$$
where $\beta$ is the known discount factor. Hence
$$
\Pr(A_1=1\mid X_1)
=
\Lambda\left[
\Delta(X_1)^\top\theta_0
+
\beta\{f_1(X_1,1)-f_1(X_1,0)\}
\right].
$$
We use the logit score
$
m_1(V,\tilde{f}_1,\theta)
=
\Delta(X_1)
\left[
A_1-\Lambda\{u(V,\tilde{f}_1,\theta)\}
\right].
$
Then the structural parameter satisfies the nonlinear vector moment equation
$
\E_0\{m_1(V,f_1,\theta_0)\}=0.
$

We use the logit score rather than the nonlinear least-squares score because its derivative is simpler. Both scores identify the same structural parameter under the model, but the logit score gives a simpler orthogonalization formula.

\subsection{Matching symbols}\label{sec:match}

To begin, we match the dynamic discrete choice model to Definition~\ref{def:nonlinear}. 
There are 
$
T=2$ time periods, and time-invariant measure $\E_t(\cdot)=\E(\cdot)$. 
The concatenation of variables is $V=(X_1,A_1,X_2,A_2)$.   The outcome is $Y=A_2$. 
For period $t=2$, we have the conditioning variables $H_2=X_2$, and the regression $f_2(X_2)=\E(A_2|X_2)$. The nonlinear formula applied to $f_2$ is $m_2(V,f_2)=Q\{f_2(X_2)\}$. 
For period $t=1$, we have the conditioning variables $H_1=(X_1,A_1)$, and the recursive regression 
$
f_1(X_1,A_1)=\E[Q\{f_2(X_2)\}|X_1,A_1]
$. The nonlinear formula applied to $f_1$ is
$
m_1(V,f_1,\theta)
=
\Delta(X_1)
\left[
A_1-\Lambda\{u(V,f_1,\theta)\}
\right].
$
Finally, the parameter of interest is identified by $
\E\{m_1(V,f_1,\theta_0)\}=0.
$

Next, we verify Assumption~\ref{assumption:nonlinear-existence}. 
For period $t=1$, we require existence of the derivative
$$
\left.
\frac{d}{d\kappa}
\E_0\{m_1(V,f_1+\kappa\tilde{f}_1,\theta_0)\}
\right|_{\kappa=0}=-\beta
\E\left[
\Delta(X_1)\Lambda'\{u(V,f_1,\theta_0)\}
\{\tilde{f}_1(X_1,1)-\tilde{f}_1(X_1,0)\}
\right],
$$
and for $a_1$ to be its Riesz representer. By the law of iterated expectations,
$$
a_1(X_1,A_1)
=
-\beta
\Delta(X_1)\Lambda'\{u(V,f_1,\theta_0)\}
\left\{
\frac{A_1}{\E(A_1 \mid X_1)}
-
\frac{1-A_1}{1-\E(A_1 \mid X_1)}
\right\},
$$
when $\E(A_1|X_1)$ is bounded away from zero and one.
For period $t=2$, we require existence of
$$
\left.
\frac{d}{d\kappa}
\E_1\{a_1(X_1,A_1)m_2(V,f_2+\kappa\tilde{f}_2)\}
\right|_{\kappa=0}=\E[a_1(X_1,A_1)Q'\{f_2(X_2)\}\tilde{f}_2(X_2)],
$$
and for $a_2$ to be its Riesz representer. Again, by the law of iterated expectations,
$$
a_2(X_2)
=
Q'\{f_2(X_2)\}\E\{a_1(X_1,A_1)\mid X_2\},
$$
when both $\E(A_1|X_1)$  and $\E(A_2|X_2)$ are bounded away from zero and one. In summary, standard overlap conditions along with bounded variances of payoffs verify Assumption~\ref{assumption:nonlinear-existence}.

Finally, we apply Theorem~\ref{theorem:nonlinear-orthogonality}. The orthogonal estimating equation is, automatically, 
$$
0
=
\E\{m_1(V,f_1,\theta_0)\}
+
\E\left(
a_1(X_1,A_1)
[Q\{f_2(X_2)\}-f_1(X_1,A_1)]
\right)
+
\E\left[
a_2(X_2)\{A_2-f_2(X_2)\}
\right].
$$
With nonlinearity, our main message remains the same: regressions recurse forward while Riesz representers recurse backward. Our automatic construction avoids analytic derivation and manual estimation of $a_1$ and $a_2$.

\subsection{Further generalizations}\label{sec:extensions}

The class of nonlinear recursive functionals in Definition~\ref{def:nonlinear} is rich enough for the dynamic discrete choice example. In the remainder of this section, we discuss three additional generalizations that may be useful for structural econometrics: (i) more moments than parameters; (ii) vector-valued regressions; and (iii) vector-valued generalized regressions. Whereas (i) and (ii) are direct extensions of Theorem~\ref{theorem:nonlinear-orthogonality}, (iii) changes the recursive conditional-mean equations themselves and requires a weighted Riesz construction.  We state the corresponding orthogonality argument but do not extend the inference theory.

\begin{remark}[More moments than parameters]\label{remark:vec1}
     So far, we have considered a vector-valued target parameter $\theta_0$ exactly identified by a vector-valued moment $m_1(V,f_1,\theta_0)\in \mathbb{R}^{J_0}$, i.e. $\dim(\theta_0)=J_0$. More generally, we may have overidentification, i.e. $\dim(\theta_0)<J_0$. Subject to the usual full-column-rank condition on the Jacobian with respect to $\theta_0$, the resulting orthogonal moment vector may be combined using the generalized method of moments.
\end{remark}

\begin{remark}[Vector-valued regressions]\label{remark:vec2}
    So far, we have considered a vector-valued target parameter $\theta_0$ identified by a vector-valued moment $m_1(V,f_t,\theta_0)\in \mathbb{R}^{J_0}$ and scalar-valued regressions. More generally, we may consider vector-valued regressions $f_t(H_t)\in \R^{J_t}$ where
    $$
  f_{t,j}(H_{t,j})
=
\E_{t,j}\{m_{t+1,j}(V,f_{t+1})\mid H_{t,j}\},\quad j=1,...,J_t,\quad t=1,...,T
    $$
    using the convention $m_{T+1,j}(V,f_{T+1})=Y_j$. 

    Let
$
a_{t,j}(H_{t,j})\in\R^{J_0}
$
denote the Riesz representer associated with the component $f_{t,j}$. The recursive debiasing term at time $t$ becomes
$
\sum_{j=1}^{J_t}
\E_{t,j}\left[
a_{t,j}(H_{t,j})
\{m_{t+1,j}(V,f_{t+1})-f_{t,j}(H_{t,j})\}
\right].
$
Assumption~\ref{assumption:nonlinear-existence} and Theorem~\ref{theorem:nonlinear-orthogonality} then apply componentwise.

If all components at time $t$ share a common history $H_t$ and measure $\E_t$, the representers may be stacked as
$
a_t(H_t)
=
\{a_{t,1}(H_t),\ldots,a_{t,J_t}(H_t)\}
\in\R^{J_0\times J_t}.
$
Writing
$
m_{t+1}(V,f_{t+1})
=
\{m_{t+1,1}(V,f_{t+1}),\ldots,
m_{t+1,J_t}(V,f_{t+1})\}^\top,
$
the recursive correction can then be written compactly as
$
\E_t\left[
a_t(H_t)
\{m_{t+1}(V,f_{t+1})-f_t(H_t)\}
\right].
$
\end{remark}

\begin{remark}[Vector-valued generalized regressions]\label{remark:vec3}
    Remark~\ref{remark:vec2} defines each component of $f_t$ as a conditional mean. More generally, each component may be a generalized regression characterized by the conditional score equation
$$
\E_{t,j}\left[
\rho_{t,j}
\left\{
m_{t+1,j}(V,f_{t+1}),
f_{t,j}(H_{t,j})
\right\}
\mid H_{t,j}
\right]
=
0,
\quad
j=1,\ldots,J_t,\quad  t=1,...,T
$$
where $\rho_{t,j}$ is a generalized residual and $m_{T+1,j}(V,f_{T+1})=Y_j$. Conditional means correspond to
$
\rho_{t,j}(z,q)=z-q,
$
while conditional $\tau_{t,j}$-quantiles correspond to
$
\rho_{t,j}(z,q)=\tau_{t,j}-1(z\leq q).
$

Define the conditional derivative
$$
v_{t,j}(H_{t,j})
=
\left.
\frac{\partial}{\partial q}
\E_{t,j}\left[
\rho_{t,j}
\left\{
m_{t+1,j}(V,f_{t+1}),
q
\right\}
\mid H_{t,j}
\right]
\right|_{q=f_{t,j}(H_{t,j})}.
$$
As in Assumption~\ref{assumption:nonlinear-existence}, let
$
\dot m_{t,j}(\tilde f_{t,j})\in\R^{J_0}
$
denote the pathwise derivative of the moment through time $t-1$ with respect to $f_{t,j}$ in the direction $\tilde f_{t,j}$. Recursive orthogonality is obtained by choosing a vector-valued Riesz representer
$
a_{t,j}(H_{t,j})\in\R^{J_0}
$
satisfying
$$
\dot m_{t,j}(\tilde f_{t,j})
+
\E_{t,j}\left[
a_{t,j}(H_{t,j})
v_{t,j}(H_{t,j})
\tilde f_{t,j}(H_{t,j})
\right]
=
0
$$
for every placeholder function $\tilde f_{t,j}(H_{t,j})$, and adding the correction
$$
\sum_{j=1}^{J_t}
\E_{t,j}\left[
a_{t,j}(H_{t,j})
\rho_{t,j}
\left\{
m_{t+1,j}(V,f_{t+1}),
f_{t,j}(H_{t,j})
\right\}
\right].
$$

When all components share a common history and measure, stacking
$
a_{t,1},\ldots,a_{t,J_t}
$
gives the matrix-valued representer
$
a_t(H_t)\in\R^{J_0\times J_t}
$
described in Remark~\ref{remark:vec2}.

For conditional means, $v_{t,j}(H_{t,j})=-1$, so the construction reduces to Assumption~\ref{assumption:nonlinear-existence}. For conditional quantiles, $-v_{t,j}(H_{t,j})$ is the conditional density of the pseudo-outcome $m_{t+1,j}(V,f_{t+1})$ evaluated at its conditional quantile. The resulting estimating equation is first-order Neyman orthogonal, although the exact mixed-bias identity for multilinear conditional-mean functionals generally does not extend without additional structure. Inference therefore requires corresponding smoothness, conditional-density, and first-stage rate conditions.
\end{remark}

\subsection{Proof of Theorem~\ref{theorem:nonlinear-orthogonality}}

Recall that
$$
\mathcal{M}(\tilde{f}_1,\ldots,\tilde{f}_T,\tilde{a}_1,\ldots,\tilde{a}_T,\theta)\\
=
\E_0\{m_1(V,\tilde{f}_1,\theta)\}
+
\sum_{t=1}^T
\E_t\left[
\tilde{a}_t(H_t)
\{m_{t+1}(V,\tilde{f}_{t+1})-\tilde{f}_t(H_t)\}
\right],
$$
with the convention $m_{T+1}(V,\tilde{f}_{T+1})=Y$. The proof exactly generalizes that of Theorem~\ref{theorem:orthogonality}.

\begin{enumerate}
    \item We verify that this moment identifies $\theta_0$ at the true nuisance functions. By an identical argument to that of Theorem~\ref{theorem:orthogonality}, all recursive debiasing terms have mean zero, so
$$
\mathcal{M}(f_1,\ldots,f_T,a_1,\ldots,a_T,\theta_0)
=
\E_0\{m_1(V,f_1,\theta_0)\}
=
0
$$
by Definition~\ref{def:nonlinear}.
    \item To verify Neyman orthogonality, we show that the pathwise derivative of $\mathcal{M}$ with respect to each nuisance functions is zero at the truth.
    \begin{enumerate}
        \item First fix a placeholder function $\tilde{a}_t(H_t)$ and define the path
$
a_t^{(\kappa)}
=
a_t+\kappa(\tilde{a}_t-a_t).
$
By an identical argument to that of Theorem~\ref{theorem:orthogonality}, for $t=1,...,T$, the pathwise derivative is
$$
\left.
\frac{d}{d\kappa}
\mathcal{M}(f_1,\ldots,f_T,a_1,\ldots,a_t^{(\kappa)},\ldots,a_T,\theta_0)
\right|_{\kappa=0}
=0.
$$
        \item Next fix a placeholder function $\tilde{f}_t(H_t)$ and define the path
$
f_t^{(\kappa)}
=
f_t+\kappa(\tilde{f}_t-f_t).
$ At the truth, $f_t$ appears only in two places: $m_t(V,f_t)$ in the previous term, and $-f_t(H_t)$ in the current term. 

For $t=1$, the pathwise derivative is
\begin{align*}
&\left.
\frac{d}{d\kappa}
\mathcal{M}(f_1^{(\kappa)},f_2,\ldots,f_T,a_1,\ldots,a_T,\theta_0)
\right|_{\kappa=0}\\
&=
\left.
\frac{d}{d\kappa}
\E_0\{m_1(V,f_1+\kappa(\tilde{f}_1-f_1),\theta_0)\}
\right|_{\kappa=0}
-
\E_1\{a_1(H_1)(\tilde{f}_1-f_1)(H_1)\}=0,
\end{align*}
where the second equality appeals to Assumption~\ref{assumption:nonlinear-existence}.

For $t=2,\ldots,T$, the pathwise derivative is
\begin{align*}
&\left.
\frac{d}{d\kappa}
\mathcal{M}(f_1,\ldots,f_t^{(\kappa)},\ldots,f_T,a_1,\ldots,a_T,\theta_0)
\right|_{\kappa=0}\\
&=
\left.
\frac{d}{d\kappa}
\E_{t-1}\{a_{t-1}(H_{t-1})m_t(V,f_t+\kappa(\tilde{f}_t-f_t))\}
\right|_{\kappa=0}
-
\E_t\{a_t(H_t)(\tilde{f}_t-f_t)(H_t)\}=0,
\end{align*}
where again second equality appeals to Assumption~\ref{assumption:nonlinear-existence}.  \qed 
    \end{enumerate}
\end{enumerate}

\section{Extension to instrumental variables}\label{sec:instrument}

The main text studies models with dynamics-on-observables, where the parameter of interest is a functional of a recursive regression. In this appendix, we generalize from recursive regressions to recursive instrumental variable functions. For clarity, we extend the class of Definition~\ref{def:linear}. Extensions of the richer classes in Appendix~\ref{sec:nonlinear}
are possible, but would add notation without changing the basic orthogonalization argument. We defer proofs to the end.

\subsection{A generalized class}

\begin{definition}[Recursive functional with instruments]\label{def:iv}
Let $V$ concatenate all observed variables across time periods, and let $T\geq1$ be fixed.
For each time period $t$, let $H_t$ denote the regressor history, let $Z_t$ denote the
instrument history, and let $\E_t(\cdot)$ denote the expectation with respect to the time-varying measure. Let
$m_t(V,f_t)$ be linear in $f_t$. We study 
$$
\theta_0=\E_0\{m_1(V,f_1)\}, \quad \E_t\{m_{t+1}(V,f_{t+1})-f_t(H_t)| Z_t\}=0, \quad t=1,\ldots,T-1,\quad 
\E_T\{Y-f_T(H_T)| Z_T\}=0,
$$
where $f_1,\ldots,f_T$ are taken to be unique functions with finite variance.
\end{definition}

Uniqueness may be guaranteed by suitable completeness conditions. If $Z_t=H_t$ for every
$t$, then Definition~\ref{def:iv} reduces to Definition~\ref{def:linear}. Now, we require two objects at each time period. 

\begin{assumption}[Existence with instruments]\label{assumption:iv-existence}
For each time period $t=1,\ldots,T$, there exist a regressor-space representer $a_t(H_t)$ and an instrument-space representer $b_t(Z_t)$, each with finite variance, 
satisfying
$$
\E_t\{a_t(H_t)\tilde f_t(H_t)\}
=
\E_{t-1}\{b_{t-1}(Z_{t-1})m_t(V,\tilde f_t)\},\quad a_t(H_t)=\E_t\{b_t(Z_t)| H_t\}
$$
for all placeholder functions $\tilde f_t(H_t)$ with finite variance. 
As a convention, $Z_0=\varnothing$ and 
$
b_0(Z_0)=1.
$
\end{assumption}

By the law of iterated expectations, Assumption~\ref{assumption:iv-existence} implies
$$
\E_t\{b_t(Z_t)\tilde f_t(H_t)\}
=
\E_t\{a_t(H_t)\tilde f_t(H_t)\}
=
\E_{t-1}\{b_{t-1}(Z_{t-1})m_t(V,\tilde f_t)\}.
$$
This identity is the instrumental variable analogue of 
Assumption~\ref{assumption:existence}.

\begin{theorem}[Recursive orthogonality with instruments]
\label{theorem:iv-orthogonality}
Consider the notation of Definition~\ref{def:iv}. Suppose
Assumption~\ref{assumption:iv-existence} holds. Define
$$
\mathcal M
(\tilde f_1,\ldots,\tilde f_T,\tilde{b}_1,\ldots,\tilde{b}_T)=
\E_0\{m_1(V,\tilde f_1)\}
+
\sum_{t=1}^T
\E_t\left[
\tilde{b}_t(Z_t)
\{m_{t+1}(V,\tilde f_{t+1})-\tilde f_t(H_t)\}
\right],
$$
with the convention $
m_{T+1}(V,f_{T+1})=Y.
$. Then $\theta_0=\mathcal{M}(\tilde f_1,\ldots,\tilde f_T,\tilde{b}_1,\ldots,\tilde{b}_T)$ is a Neyman orthogonal estimating equation. Here, $f_t$ is recursively defined from $f_{t+1}$ in Definition~\ref{def:iv}, and $b_t$ is recursively defined from $b_{t-1}$ below Assumption~\ref{assumption:iv-existence}.
\end{theorem}

The additional difficulty in the instrumental variable setting can be seen from operator notation. Define
the conditional expectation operator
$
(\mathcal T_t \tilde{f}_t)(Z_t)
=
\E_t\{\tilde{f}_t(H_t)| Z_t\},
$
and its adjoint
$
(\mathcal T_t^*\tilde{b}_t)(H_t)
=
\E_t\{\tilde{b}_t(Z_t)| H_t\}.
$
The instrumental variable function satisfies
$
(\mathcal T_t f_t)(Z_t)
=
\E_t\{m_{t+1}(V,f_{t+1})| Z_t\},
$
while the instrument-space representer satisfies
$
(\mathcal T_t^*b_t)(H_t)=a_t(H_t).
$
Thus each time period contains both a primal and dual inverse problem. When
$Z_t=H_t$, both operators reduce to the identity and $b_t=a_t$, recovering
Theorem~\ref{theorem:orthogonality}.

With instruments, estimation may be ill posed. Across several time periods, the
ill-posedness of the inverse problems may compound. Mean-square rates and inference for recursive inverse problems with
generic machine learning require additional source, completeness, and
well-posedness conditions \citep{meza2021nested}.

\subsection{Proof of Theorem~\ref{theorem:iv-orthogonality}}
Recall that
$$
\mathcal M
(\tilde f_1,\ldots,\tilde f_T,\tilde{b}_1,\ldots,\tilde{b}_T)=
\E_0\{m_1(V,\tilde f_1)\}
+
\sum_{t=1}^T
\E_t\left[
\tilde{b}_t(Z_t)
\{m_{t+1}(V,\tilde f_{t+1})-\tilde f_t(H_t)\}
\right]
$$
with the convention $m_{T+1}(V,\tilde{f}_{T+1})=Y$. The proof exactly generalizes that of Theorem~\ref{theorem:orthogonality}.

\begin{enumerate}
    \item We verify that the moment identifies $\theta_0$ at the true nuisance functions. By the law of iterations and the recursive definition of $f_t$, for each $t=1,...,T$
$$
\E_t\left[
b_t(Z_t)
\{m_{t+1}(V,f_{t+1})-f_t(H_t)\}
\right]=
\E_t\left[
b_t(Z_t)
\E_t\{m_{t+1}(V,f_{t+1})-f_t(H_t)| Z_t\}
\right]=0.
$$
Hence all recursive corrections have mean zero, so
$$
\mathcal M
(f_1,\ldots,f_T,b_1,\ldots,b_T)
=
\E_0\{m_1(V,f_1)\}
=
\theta_0.
$$
    \item To verify Neyman orthogonality, we show that the pathwise derivative of $\mathcal{M}$ with respect  each nuisance function is zero at the truth.

\begin{enumerate}
    \item First fix a placeholder function $\tilde{b}_t(Z_t)$ and define the path
$
b_t^{(\kappa)}
=
b_t+\kappa(\tilde{b}_t-b_t).
$
At the truth, only the $t$-th recursive debiasing term depends on $b_t$. Therefore, for $t=1,...,T,$ the pathwise derivative is
\begin{align*}
&\left.
\frac{d}{d\kappa}
\mathcal M
(f_1,\ldots,f_T,
b_1,\ldots,b_t^{(\kappa)},\ldots,b_T)
\right|_{\kappa=0}=
\E_t\left[
\{\tilde{b}_t(Z_t)-b_t(Z_t)\}
\{m_{t+1}(V,f_{t+1})-f_t(H_t)\}
\right]\\
&\quad=
\E_t\left[
\{\tilde{b}_t(Z_t)-b_t(Z_t)\}
\E_t\{m_{t+1}(V,f_{t+1})-f_t(H_t)| Z_t\}
\right]
=0
\end{align*}
by scalar differentiation, the law of iterated expectations, and the recursive definition of $f_t$.
    \item Next fix a placeholder function $\tilde f_t(H_t)$ and define the path
$
f_t^{(\kappa)}
=
f_t+\kappa(\tilde f_t-f_t).
$
At the truth, $f_t$ appears only in two places: $m_t(V,f_t)$ in the
previous term term, and $-f_t(H_t)$ in the current term. Recall that $f_t\mapsto m_t(V,f_t)$ is linear by Definition~\ref{def:iv}.

For $t=1$, the pathwise derivative is
\begin{align*}
&\left.
\frac{d}{d\kappa}
\mathcal M
(f_1^{(\kappa)},f_2,\ldots,f_T,b_1,\ldots,b_T)
\right|_{\kappa=0}
=
\E_0\{m_1(V,\tilde f_1-f_1)\}
-
\E_1\{b_1(Z_1)(\tilde f_1-f_1)(H_1)\}\\
&\quad=
\E_0\{m_1(V,\tilde f_1-f_1)\}
-
\E_1\{a_1(H_1)(\tilde f_1-f_1)(H_1)\}=0,
\end{align*}
where the second equality uses
$
\E_1\{b_1(Z_1)| H_1\}=a_1(H_1),
$
and the final equality uses the Riesz equation in
Assumption~\ref{assumption:iv-existence}.

For $t=2,\ldots,T$, the same argument gives
\begin{align*}
&\left.
\frac{d}{d\kappa}
\mathcal M
(f_1,\ldots,f_t^{(\kappa)},\ldots,f_T,b_1,\ldots,b_T)
\right|_{\kappa=0}\\
&\quad=
\E_{t-1}\{b_{t-1}(Z_{t-1})m_t(V,\tilde f_t-f_t)\}
-
\E_t\{b_t(Z_t)(\tilde f_t-f_t)(H_t)\}\\
&\quad=
\E_{t-1}\{b_{t-1}(Z_{t-1})m_t(V,\tilde f_t-f_t)\}
-
\E_t\{a_t(H_t)(\tilde f_t-f_t)(H_t)\}
=0,
\end{align*}
again by the two equations in Assumption~\ref{assumption:iv-existence}.
\end{enumerate}
\end{enumerate}

\section{Connection to clever covariates}\label{sec:clever}

In the main text, Algorithm~\ref{algo:dml} adds recursive debiasing terms to the identifying formula. We now describe an alternative implementation that absorbs the explicit debiasing terms into targeted updates of the recursive regressions. In doing so, we formalize the connection between our nonparametric method and the parametric clever-covariate method of \citet{bang2005doubly}.

The following procedure is a recursive Riesz analogue of clever-covariate adjustment. After estimating a preliminary regression $\hat f_t$ and Riesz representer $\hat a_t$, consider the targeted update
$$
\hat f_t^{\operatorname{clever}}(H_t)
=
\hat f_t(H_t)
+
\hat\epsilon_t\hat a_t(H_t).
$$
Working backward from $t=T$ to $t=1$, choose $\hat\epsilon_t$ by
$$
\hat\epsilon_t
\in
\argmin_{\epsilon\in\R}
\hat{\E}_t\left[
\left\{
m_{t+1}(V,\hat f_{t+1}^{\operatorname{clever}})
-
\hat f_t(H_t)
-
\epsilon\cdot \hat a_t(H_t)
\right\}^2
\right],
$$
using the convention
$
m_{T+1}(V,\hat f_{T+1}^{\operatorname{clever}})=Y.
$
Equivalently, the pseudo-outcome
$
m_{t+1}(V,\hat f_{t+1}^{\operatorname{clever}})
$
is regressed on the clever covariate $\hat a_t(H_t)$, with the preliminary regression $\hat f_t(H_t)$ used as an offset.

With clever covariate adjustments applied to the recursive regressions, our debiased estimator in Algorithm~\ref{algo:dml} simplifies to a plug-in estimator. Formally, because the scalar coefficient $\hat\epsilon_t$ is unpenalized, its first-order condition is
$$
\hat{\E}_t\left[
\hat a_t(H_t)
\left\{
m_{t+1}(V,\hat f_{t+1}^{\operatorname{clever}})
-
\hat f_t^{\operatorname{clever}}(H_t)
\right\}
\right]
=
0.
$$
Thus, on the sample used for the targeted update, the recursive debiasing term at time $t$ is zero. Applying these updates backward from $t=T$ to $t=1$ sets every recursive debiasing term to zero. The resulting estimator can therefore be written in the plug-in form
$$
\hat\theta^{\operatorname{clever}}
=
\hat{\E}_0\{m_1(V,\hat f_1^{\operatorname{clever}})\}.
$$

Sample splitting and cross-fitting require fold-specific targeted updates. For each evaluation fold, the preliminary regressions and Riesz representers are estimated on its complement, while the scalar targeting coefficients are estimated on the evaluation fold so that the corresponding empirical first-order conditions hold there. The resulting fold-specific plug-in estimates may then be averaged across folds.

We do not formally analyze this alternative implementation. Algorithm~\ref{algo:dml} avoids the additional targeting step by retaining the recursive debiasing terms explicitly.

\section{Simulation details}\label{sec:sim}

\subsection{Example~\ref{ex:time}}

\begin{table}
\centering
\caption{Linear time-varying treatment effect simulation: Additional estimators}
\label{tab:time_appendix_linear}
\small
\setstretch{1.0}
\setlength{\tabcolsep}{10pt}
\begin{tabular}{clcccc}
\toprule
\toprule
$n$ & Method & Bias & RMSE & Coverage & CI Length \\
\midrule
\multirow{14}{*}{500}
 & Oracle                              &   0.023 &   0.245 & 0.932 &   0.877 \\
 & Manual-Lasso-Lasso (Sequential)     &   0.076 &   0.611 & 0.942 &   1.295 \\
 & Manual-Lasso-Ridge (Sequential)     &  -0.318 &   1.226 & 0.964 &   2.566 \\
 & Manual-RF-Ridge (Sequential)        &   1.280 &   4.644 & 0.946 &   6.587 \\
 & Manual-NN-Ridge (Sequential)        &   0.337 &   2.578 & 0.978 &   3.947 \\
 & Manual-Lasso-Lasso                  &  -0.235 &   1.020 & 0.962 &   2.270 \\
 & Manual-RF-RF                        &  -0.678 &   0.732 & 0.216 &   0.917 \\
 & Manual-NN-NN                        &  31.262 & 209.705 & 0.962 & 143.069 \\
 & Manual-Lasso-Ridge                  &  -0.352 &   1.225 & 0.968 &   2.551 \\
 & Manual-RF-Ridge                     &   1.360 &   4.672 & 0.952 &   6.641 \\
 & Manual-NN-Ridge                     &   0.337 &   2.578 & 0.978 &   3.947 \\
 & Auto-Lasso-Lasso                         &   0.023 &   0.229 & 0.928 &   0.815 \\
 & Auto-RF-RF                             &   0.019 &   0.237 & 0.858 &   0.711 \\
 & Auto-NN-NN                             &  -0.238 &   0.333 & 0.744 &   0.799 \\
\cmidrule(lr){1-6}
\multirow{14}{*}{1000}
 & Oracle                              &   0.000 &   0.162 & 0.942 &   0.623 \\
 & Manual-Lasso-Lasso (Sequential)     &   0.012 &   0.248 & 0.942 &   0.837 \\
 & Manual-Lasso-Ridge (Sequential)     &  -0.259 &   0.844 & 0.978 &   1.861 \\
 & Manual-RF-Ridge (Sequential)        &   0.629 &   2.609 & 0.956 &   3.312 \\
 & Manual-NN-Ridge (Sequential)        &   0.357 &   1.698 & 0.978 &   2.891 \\
 & Manual-Lasso-Lasso                  &  -0.223 &   0.754 & 0.976 &   1.745 \\
 & Manual-RF-RF                        &  -0.246 &   0.302 & 0.570 &   0.554 \\
 & Manual-NN-NN                        & 109.190 & 1926.397 & 0.976 & 433.216 \\
 & Manual-Lasso-Ridge                  &  -0.292 &   0.876 & 0.974 &   1.855 \\
 & Manual-RF-Ridge                     &   0.652 &   2.623 & 0.956 &   3.325 \\
 & Manual-NN-Ridge                     &   0.357 &   1.698 & 0.978 &   2.891 \\
 & Auto-Lasso-Lasso                          &  -0.003 &   0.149 & 0.938 &   0.564 \\
 & Auto-RF-RF                             &   0.001 &   0.156 & 0.894 &   0.506 \\
 & Auto-NN-NN                             &  -0.039 &   0.177 & 0.906 &   0.637 \\
\cmidrule(lr){1-6}
\multirow{14}{*}{2000}
 & Oracle                              &   0.005 &   0.110 & 0.970 &   0.441 \\
 & Manual-Lasso-Lasso (Sequential)     &  -0.003 &   0.238 & 0.976 &   0.650 \\
 & Manual-Lasso-Ridge (Sequential)     &  -0.264 &   1.107 & 0.984 &   1.523 \\
 & Manual-RF-Ridge (Sequential)        &   0.420 &   1.734 & 0.974 &   2.117 \\
 & Manual-NN-Ridge (Sequential)        &   0.344 &   1.412 & 0.984 &   2.175 \\
 & Manual-Lasso-Lasso                  &  -0.244 &   1.042 & 0.984 &   1.465 \\
 & Manual-RF-RF                        &  -0.172 &   0.210 & 0.560 &   0.381 \\
 & Manual-NN-NN                        &  22.556 & 213.570 & 0.986 &  92.942 \\
 & Manual-Lasso-Ridge                  &  -0.290 &   1.122 & 0.982 &   1.529 \\
 & Manual-RF-Ridge                     &   0.444 &   1.741 & 0.974 &   2.128 \\
 & Manual-NN-Ridge                     &   0.344 &   1.412 & 0.984 &   2.175 \\
 & Auto-Lasso-Lasso                          &   0.008 &   0.100 & 0.958 &   0.397 \\
 & Auto-RF-RF                             &   0.005 &   0.135 & 0.906 &   0.368 \\
 & Auto-NN-NN                             &   0.002 &   0.148 & 0.962 &   0.564 \\
\bottomrule
\bottomrule
\end{tabular}
\end{table}

\begin{table}
\centering
\caption{Nonlinear time-varying treatment effect simulation: Additional estimators}
\label{tab:time_appendix_nonlinear}
\small
\setstretch{1.0}
\setlength{\tabcolsep}{10pt}
\begin{tabular}{clcccc}
\toprule
\toprule
$n$ & Method & Bias & RMSE & Coverage & CI Length \\
\midrule
\multirow{14}{*}{500}
 & Oracle                              &  -0.013 &   0.268 & 0.956 &   1.049 \\
 & Manual-Lasso-Lasso (Sequential)     &   0.143 &   0.298 & 0.902 &   1.048 \\
 & Manual-Lasso-Ridge (Sequential)     &   0.124 &   0.292 & 0.926 &   1.092 \\
 & Manual-RF-Ridge (Sequential)        &   0.033 &   0.877 & 0.924 &   2.458 \\
 & Manual-NN-Ridge (Sequential)        &   0.191 &   0.677 & 0.990 &   2.436 \\
 & Manual-Lasso-Lasso                  &   0.125 &   0.291 & 0.922 &   1.085 \\
 & Manual-RF-RF                        &   0.803 &   0.862 & 0.464 &   1.525 \\
 & Manual-NN-NN                        & -71.203 & 1480.870 & 0.978 & 282.951 \\
 & Manual-Lasso-Ridge                  &   0.122 &   0.291 & 0.926 &   1.093 \\
 & Manual-RF-Ridge                     &   0.052 &   0.866 & 0.922 &   2.462 \\
 & Manual-NN-Ridge                     &   0.191 &   0.678 & 0.990 &   2.436 \\
 & Auto-Lasso-Lasso                         &   0.171 &   0.311 & 0.910 &   1.027 \\
 & Auto-RF-RF                             &   0.080 &   0.285 & 0.930 &   1.051 \\
 & Auto-NN-NN                          &   0.234 &   0.356 & 0.822 &   0.999 \\
\cmidrule(lr){1-6}
\multirow{14}{*}{1000}
 & Oracle                              &  -0.003 &   0.197 & 0.934 &   0.739 \\
 & Manual-Lasso-Lasso (Sequential)     &   0.150 &   0.248 & 0.882 &   0.735 \\
 & Manual-Lasso-Ridge (Sequential)     &   0.143 &   0.246 & 0.884 &   0.753 \\
 & Manual-RF-Ridge (Sequential)        &   0.148 &   0.406 & 0.910 &   1.538 \\
 & Manual-NN-Ridge (Sequential)        &   0.169 &   0.353 & 0.986 &   1.608 \\
 & Manual-Lasso-Lasso                  &   0.144 &   0.246 & 0.884 &   0.751 \\
 & Manual-RF-RF                        &   0.441 &   0.507 & 0.704 &   1.128 \\
 & Manual-NN-NN                        &  -0.910 &   5.080 & 0.990 &   5.116 \\
 & Manual-Lasso-Ridge                  &   0.141 &   0.245 & 0.884 &   0.753 \\
 & Manual-RF-Ridge                     &   0.119 &   0.410 & 0.924 &   1.579 \\
 & Manual-NN-Ridge                     &   0.169 &   0.353 & 0.986 &   1.608 \\
 & Auto-Lasso-Lasso                          &   0.180 &   0.264 & 0.840 &   0.721 \\
 & Auto-RF-RF                             &   0.063 &   0.214 & 0.910 &   0.753 \\
 & Auto-NN-NN                             &   0.073 &   0.217 & 0.928 &   0.762 \\
\cmidrule(lr){1-6}
\multirow{14}{*}{2000}
 & Oracle                              &  -0.016 &   0.132 & 0.954 &   0.524 \\
 & Manual-Lasso-Lasso (Sequential)     &   0.138 &   0.189 & 0.818 &   0.519 \\
 & Manual-Lasso-Ridge (Sequential)     &   0.133 &   0.186 & 0.830 &   0.527 \\
 & Manual-RF-Ridge (Sequential)        &   0.191 &   0.304 & 0.846 &   1.016 \\
 & Manual-NN-Ridge (Sequential)        &   0.133 &   0.247 & 0.980 &   1.126 \\
 & Manual-Lasso-Lasso                  &   0.133 &   0.186 & 0.830 &   0.526 \\
 & Manual-RF-RF                        &   0.347 &   0.386 & 0.600 &   0.783 \\
 & Manual-NN-NN                        &  -0.197 &   0.809 & 1.000 &   1.743 \\
 & Manual-Lasso-Ridge                  &   0.131 &   0.185 & 0.836 &   0.527 \\
 & Manual-RF-Ridge                     &   0.101 &   0.272 & 0.920 &   1.052 \\
 & Manual-NN-Ridge                     &   0.134 &   0.247 & 0.980 &   1.127 \\
 & Auto-Lasso-Lasso                          &   0.171 &   0.214 & 0.748 &   0.512 \\
 & Auto-RF-RF                             &   0.028 &   0.138 & 0.952 &   0.541 \\
 & Auto-NN-NN                             &   0.063 &   0.151 & 0.914 &   0.544 \\
\bottomrule
\bottomrule
\end{tabular}
\end{table}

\textbf{Additional estimators.} Tables~\ref{tab:time_main_linear} and~\ref{tab:time_main_nonlinear} in the main text focus on one manual debiased machine learning estimator for brevity. Tables~\ref{tab:time_appendix_linear} and~\ref{tab:time_appendix_nonlinear} augment Tables~\ref{tab:time_main_linear} and~\ref{tab:time_main_nonlinear} with several additional variations of manual debiased machine learning. The takeaways remain the same: our method outperforms all variations of manual debiased machine learning.

The four variations of manual debiased machine learning are as follows. 

(i) Manual-Lasso-Lasso (Sequential) is what was previously called Manual-Lasso in Tables~\ref{tab:time_main_linear} and~\ref{tab:time_main_nonlinear}. It uses lasso for regression and propensity models, and the doubly robust moment with debiased pseudo outcomes.

(ii) Manual-Lasso-Ridge (Sequential) uses lasso for the regression models, ridge for the propensity models, and the doubly robust moment with debiased pseudo outcomes. Using ridge propensity scores is an attempt to alleviate numerical instability. Manual-RF-Ridge (Sequential) and Manual-NN-Ridge (Sequential) are similar, instead using random forests or neural networks, respectively, for the regression models.

(iii) Manual-Lasso-Lasso uses lasso for regression and propensity models, and the doubly robust moment with plug-in pseudo outcomes. Manual-RF-RF and Manual-NN-NN are similar, instead using random forests or neural networks, respectively, for the regression models.

(iv) Finally, Manual-Lasso-Ridge is like Manual-Lasso-Lasso, but with ridge propensity models. As before, using ridge propensity scores is an attempt to alleviate numerical instability.  Manual-RF-Ridge and Manual-RF-Ridge are similar.

\textbf{Simulation design.} Both the linear and nonlinear designs share a common structure, based on \cite{bradic2024high}. Given the covariate dimension $d_X$, link function $\pi(\cdot)$, link coefficients $\beta_1\in\mathbb{R}^{d_X+1}$ and $\beta_2^{(0)},\beta_2^{(1)}\in\mathbb{R}^{2d_X+1}$, and regression $f_2$, each draw of $V=(X_1,D_1,X_2,D_2,Y)$ is generated as follows:
\begin{align*}
    &X_1\sim \mathcal{N}(0,I_{d_X}), \quad D_1|X_1\sim \operatorname{Bernoulli}[\pi\{(1,X_1^{\top})^{\top}\beta_1\}] \\
    &X_2=X_1+\eta_1D_1 1_{d_X}+\eta_2,\quad \eta_1\sim \mathcal{N}(1,1),\quad \eta_2\sim\mathcal{N}(0,I_{d_X}) \\
    & D_2|X_1,D_1,X_2 \sim \operatorname{Bernoulli}[\pi\{(1,X_1^{\top},X_2^{\top})^{\top}\beta_2^{(D_1)}\}] \\
    &Y=f_2(X_1,D_1,X_2,D_2)+\varepsilon,\quad \varepsilon\sim\mathcal{N}(0,1).
\end{align*}
Here, $I_{d_X}\in\mathbb{R}^{d_X\times d_X}$ is the identity matrix and  $1_{d_X}\in\mathbb{R}^{d_X}$ is the vector of ones. The target parameter $\theta_0$ is the mean potential outcome under the treatment sequence $(d_1,d_2)=(1,1)$.

For the linear design, the covariate dimension is $d_X=5$. The link function is the truncated logistic $\pi(s)=0.1+(0.9-0.1)\frac{\exp(s)}{1+\exp(s)}$ with link coefficients 
$$
\beta_1=(0,1,1,1,0,0)^\top,\quad \beta_2^{(0)}
=
(0,0.5,0,-0.5,0,0,0.5,0,0.5,0,0)^\top,\quad \beta_2^{(1)}
=
(0,1,1,0,0,0,1,-1,0,0,0)^\top.
$$
The regression is
$
f_2(X_1,D_1,X_2,D_2)=(1,X_1,X_2)^{\top} \gamma^{(D_1,D_2)}
$ where
\begin{align*}
    \gamma^{(1,1)}&= (-1,-1,1,-1,0,0,-1,-1,1,0,0)^{\top}, \quad \gamma^{(1,0)}= (0,0,0,0,0,0,0,0,0,0,0)^{\top} \\
 \gamma^{(0,1)}&= (0,0,0,0,0,0,0,0,0,0,0)^{\top},\quad \gamma^{(0,0)}= (1,1,1,-1,0,0,1,1,1,0,0)^{\top}.
\end{align*}
By construction, $\theta_0=-2$.

For the nonlinear design, the covariate dimension is $d_X=2$. The link function is the step function $\pi(s)=0.3+(0.7-0.3)1(s>0)$ with link coefficients 
$$
\beta_1= (0, 1, 0)^\top,\quad \beta_2^{(0)} = (0, 0,0,0.5, 0)^\top,\quad \beta_2^{(1)} = (0,0,0, 1, 0)^\top.
$$
Using the selector notation $e_1=(1,0)^\top$, the regression is
\begin{align*}
    f_2(X_1,D_1,X_2,D_2)&=e_1^\top X_1
+
(2.2+e_1^\top X_1)D_1
+
(1.2+e_1^\top X_2)D_2
+
0.5D_1D_2 \\
&\quad + 1.2 \cdot 1(e_1^\top X_1>0) + 0.8 \cdot 1(e_1^\top X_2>0) + 0.5D_1D_2(e_1^\top X_2).
\end{align*}
By construction, $\theta_0=6.575$.

\textbf{Hyperparameter tuning.} First we desribe the subset of estimators in Tables~\ref{tab:time_main_linear} and~\ref{tab:time_main_nonlinear}.

The oracle uses no cross fitting.

Manual-Lasso is implemented using code from \cite{bradic2024high}. It has linear regression models and logistic propensity models, with $4$-fold cross fitting. The $\ell_1$ penalty is selected by cross validation.

Auto-Lasso is implemented by recursing code from \cite{chernozhukov2022automatic}. It has linear regression models and linear Riesz models, with $4$-fold cross fitting. The $\ell_1$ penalty according to the theoretical rule of \cite{belloni2014inference}.

Auto-RF is implemented by recursing code from \cite{chernozhukov2022riesznet}. It has random forest regression models and random forest Riesz models, with $4$-fold cross fitting. It uses $100$ trees, a minimum leaf size of $5$, and a subsampling fraction of $0.45$.

Auto-NN is implemented by recursing code from \cite{chernozhukov2022riesznet}. It has neural network regression models and neural network Riesz models, with $4$-fold cross fitting. Across models there are $200$ shared hidden units, $100$ task-specific hidden units, exponential linear activation, Adam optimization, learning rate of $10^{-4}$, weight decay of $10^{-3}$, and early stopping after $20$ rounds without improvement of at least $10^{-3}$.

Next we describe the additional estimators in Tables~\ref{tab:time_appendix_linear} and~\ref{tab:time_appendix_nonlinear}. These are implemented with the code of \cite{meza2021nested} following their default settings for lasso, random forest, or neural network regression models with lasso, random forest, neural network, or ridge-regularized logistic propensity models. We use $4$-fold cross fitting.

\subsection{Example~\ref{ex:diff}}

\textbf{Simulation design.} Each draw of $V=(Z,X_{\pre},Y_{\pre},D,X_{\post},Y_{\post})$ is generated as follows:
\begin{align*}
    &Z\sim\mathcal{N}(0,I_2),\quad X_{\pre}\sim\mathcal{N}(0,I_3),\quad Y_{\pre}=2\cdot 1(X_{\pre}^{\top}e_1>0)+X_{\pre}^{\top}e_1+Z^{\top}1_2+\varepsilon,\quad \varepsilon\sim\mathcal{N}(0,1) \\
    &D|Z,X_{\pre},Y_{\pre},\eta \sim  \operatorname{Bernoulli}[\pi\{(1,Z^{\top},X_{\pre}^{\top},Y_{\pre}^{\top})^{\top}\beta_0+\eta\}],\quad \beta_0 = (0, 1, 1, 1, 0, 0, 1)^\top \quad \eta\sim\mathcal{N}(0,1) \\
    &X_{\post}=X_{\pre}+(1+\eta)\cdot D\cdot 1_3  +\xi,\quad \xi\sim\mathcal{N}(0,I_3) \\
    &Y_{\post}=Y_{\pre}+D+2\cdot\{1(X_{\post}^{\top}e_1>0)+D\cdot (X_{\post}^{\top}e_1)\}+1(Y_{\pre}>0)+\zeta,\quad \zeta\sim\mathcal{N}(0,1).
\end{align*}
Similar to the earlier notation, $I_{\dim}\in\mathbb{R}^{\dim \times \dim}$ is an identity, $1_{\dim}\in\mathbb{R}^{\dim}$ is a vector of ones, $e_1=(1,0,0)$, and $\pi(s)=0.3+(0.7-0.3)1(s>0)$ is the step link function. By construction, the average treatment effect on the treated is $\theta_0=3.954$.

\textbf{Hyperparameter tuning.} OLS is implemented according to \cite{caetano2022difference}, with bootstrap standard errors.
Auto-Lasso is implemented using linear regression models and linear Riesz models, with 5-fold cross fitting. The $\ell_1$ penalty is selected by cross validation.

Auto-RF and Auto-NN are implemented as before, with $5$-fold cross fitting.
\FloatBarrier
\section{Application details}\label{sec:app}

\subsection{Example~\ref{ex:surrogate}}

\textbf{Data.} The dataset we use comes from unpublished replication materials accompanying \cite{athey2025surrogate}. We received the dataset directly from the authors and have permission to use it for our analysis. Researchers interested in accessing these data can contact the authors of \cite{athey2025surrogate}.

We conduct a program evaluation of the Greater Avenues to Independence (GAIN) job training program. Eligibility was randomly assigned in certain counties of California in 1988. Following the literature, we focus on four sites: Alameda, Los Angeles, Riverside, and San Diego.

The variable definitions follow \cite{athey2025surrogate}. The pre-treatment covariates $X\in\R^{37}$ are gender, education, family status, race, age, and lagged pre-treatment values of aid receipt, employment, and earnings. The treatment $D\in\{0,1\}$ is randomized eligibility for job training. The surrogate outcomes $S\in\mathbb{R}^{12}$ are quarterly earnings and program aid receipt measured for six quarters, i.e. approximately every three months for $18$ months after program entry. The long run outcome $Y\in\mathbb{R}$ is the average of quarterly earnings over $36$ quarters post-treatment, i.e. approximately over the nine years after program entry.

The sample definitions also follow \cite{athey2025surrogate}. The short-term experimental group, for which we observe $(X,D,S)$, is Riverside. The long-term observational group, for which we observe $(X,S,Y)$, consists of Alameda, Los Angeles, and San Diego.

Finally, the estimand $\theta_0$ follows \cite{athey2025surrogate}: the long run average treatment effect for the experimental subpopulation. Relative to Example~\ref{ex:surrogate} in the main text, the only difference is that here we take $\E_0(\cdot)=\E_{\exp}(\cdot)$; the derivations remain identical with an appropriate change of measure.

\textbf{Hyperparameter tuning.} The benchmark is the estimate an economist would obtain from oracle data access of $(D,Y)$ in the experimental group. It is the difference in means of $Y$ for the treated experimental and untreated experimental subgroups.

The observational estimate is what an economist would obtain from only access to the confounded data $(D,Y)$ in the observational group.  It is the difference in means of $Y$ for the treated observational and untreated observational subgroups, which suffers from selection bias.

Manual-Lasso, Manual-RF, and Manual-NN are implemented with the code of \cite{meza2021nested}. Following their default settings, we use lasso, random forest, or neural network regression models with ridge-regularized logistic propensity models and $5$-fold cross fitting.

Auto-Lasso is implemented by recursing code from \cite{chernozhukov2022automatic}, with $5$-fold cross fitting. The $\ell_1$ penalty is selected by the theoretical rule of \cite{belloni2014inference}.

Auto-RF and Auto-NN are implemented as before, with $5$-fold cross fitting. 

\subsection{Example~\ref{ex:diff}}

\textbf{Data.} The dataset we use comes from published and publicly available replication materials accompanying \cite{chernozhukov2023automatic}. 

We conduct a program evaluation of the federal minimum wage in 2004, using annual county records from 2001 through 2007. Following the literature, we place parallel trends assumptions.

The variable definitions follow \cite{callaway2021difference}. The time invariant covariates $Z\in \R^{6}$ are region, baseline employment, baseline population, and baseline average pay measured in 2001. The pre-treatment covariates $X_{\pre}\in\R^{2}$ are population and average pay measured in 2003, while the pre-treatment outcome $Y_{\pre}\in\R$ is the logarithm of teen employment measured in 2003. The treatment $D\in\{0,1\}$ indicates whether a county raised its minimum wage above the federal minimum wage in 2004. The post-treatment covariates $X_{\post}\in\mathbb{R}^{2}$ are population and average pay measured in either 2004, 2005, 2006, or 2007, corresponding to whether the post-treatment outcome $Y_{\post}\in\mathbb{R}$ is taken to be the logarithm of teen employment measured in either 2004, 2005, 2006, or 2007, respectively.

\textbf{Hyperparameter tuning.} The static model assumes parallel trends conditional upon $(Z,X_{\pre})$. 
OLS is implemented with robust standard errors. 
Manual-Linear is the doubly robust estimator of \cite{sant2020doubly}, implemented with linear regression and logistic propensity score, without cross fitting.
Auto-Lasso, Auto-RF, and Auto-NN are implemented using code from \cite{chernozhukov2023automatic}, all with $5$-fold cross fitting. For Auto-Lasso, the $\ell_1$ penalty is selected by cross validation. Auto-RF has $10$ trees, a minimum leaf size of $5$, and a subsampling fraction of $0.45$. Auto-NN is as before.

The dynamic model assumes parallel trends conditional upon $(Z,X_{\pre},X_{\post})$. The OLS method of \cite{caetano2022difference} is reported with bootstrap standard errors. Auto-Lasso, Auto-RF, and Auto-NN are implemented by recusing code from \cite{chernozhukov2023automatic}, all with $5$-fold cross fitting. For Auto-Lasso, the $\ell_1$ penalty is selected by cross validation. Auto-RF has $100$ trees, a minimum leaf size of $5$, and a subsampling fraction of $0.45$. Auto-NN is as before.

\end{document}